\documentclass[11pt,a4paper]{article}
\usepackage{jheppub_kim}

\usepackage{amsmath}
\usepackage{amssymb}
\usepackage{dcolumn}
\usepackage{bm}
\usepackage{color}
\usepackage{epsfig}
\usepackage{amsfonts}

\title{Dynamical analysis of generalized Galileon cosmology}

\author[a,b]{Genly Leon}
\author[c,d]{and Emmanuel N. Saridakis}
\affiliation[a]{Departamento de
Matem\'atica, Universidad Central de Las Villas, Santa Clara
\enskip CP 54830, Cuba}

\affiliation[b]{Instituto de F\'{\i}sica, Pontificia Universidad
Cat\'olica de
Valpara\'{\i}so,
Casilla 4950, Valpara\'{\i}so, Chile}

\affiliation[c]{Physics Division, National Technical
University of Athens, 15780 Zografou Campus,  Athens, Greece}
\affiliation[d]{CASPER, Physics Department, Baylor University,
Waco, TX  76798-7310, USA}

\emailAdd{genly.leon@ucv.cl}
\emailAdd{Emmanuel$_-$Saridakis@baylor.edu}

\abstract{We perform a detailed dynamical analysis of generalized Galileon
cosmology, incorporating also the requirements of ghost and instabilities
absence. We find that there are not any new stable late-time solutions
apart from those of standard quintessence. Furthermore, depending on the
model parameters the Galileons may survive at late times or they may
completely disappear by the dynamics, however the corresponding
observables are always independent of the  Galileon terms, determined
only by the usual action terms. Thus, although the Galileons can play an
important role at inflationary or at recent times, in the future, when the
universe will asymptotically reach its stable state, they will not have any
effect on its evolution.}
\arxivnumber{}
\begin{document}
\maketitle

\section{Introduction}

There are two approaches one can follow in order to describe the observed
universe acceleration. The first is to introduce the concept of dark
energy, usually adding extra scalar fields (see
\cite{Copeland:2006wr} and references therein) in the
right-hand-side of the field equations of General Relativity, while the
second is to modify the left-hand-side of the general relativistic field
equations, that is to modify the gravitational theory itself (see
\cite{Nojiri:2006ri,Capozziello:2011et} and references therein).

A recently re-discovered, very general class of scalar-field theories is
based on the introduction of higher derivatives in the action, with the
requirement of maintaining the equations of motion second order.
Although the most general second-order theories avoiding the
Ostrogradsky instabilities \cite{Ost} were already derived in
\cite{Horndeski}, a particular class, dubbed Galileon, was constructed in
\cite{Nicolis:2008in} for the Minkowski metric and in
\cite{Deffayet:2009wt,Deffayet:2009mn} for dynamical geometries, while in
\cite{Deffayet:2011gz} the results of \cite{Horndeski} were re-discovered
in
the context of (extended) Galileon framework. In this formalism the
Lagrangian is suitably constructed in order for the field equations to be
invariant under the Galilean symmetry $\phi\rightarrow\phi+c$,
$\partial_\mu \phi\rightarrow\partial_\mu \phi+b_\mu$ in the limit of
Minkowski spacetime, with $c,b_\mu$ constants. The four-dimensional
Lagrangian that preserves these symmetries contains five unique
terms consisting of scalar combinations of $\partial_\mu\phi$,
$\partial_\mu\partial_\nu\phi$ and $\Box\phi$ (since General Relativity
does not accept  the satisfaction of Galilean symmetry in curved spacetime,
higher derivatives are necessary in the action  \cite{Nicolis:2009qm,
Deffayet:2010qz,Germani:2011bc}), and the corresponding
couplings are not renormalized by loop corrections
\cite{Hinterbichler:2010xn}. Furthermore, a significant advantage is that
the derivative self-couplings of the scalar screen the deviations from
General Relativity at high gradient regions (small scales or high
densities) through the Vainshtein mechanism \cite{Vainshtein:1972sx},
thus satisfying solar system and early universe constraints
\cite{Chow:2009fm,DeFelice:2011th,Babichev:2011kq}.

Application of the above construction in cosmological frameworks gives
rise to the Galileon cosmology, which proves to be very interesting and
has been investigated in detail in the literature. In particular, one can
investigate the universe evolution and late-time acceleration
\cite{Silva:2009km,Gannouji:2010au,DeFelice:2010pv,Mota:2010bs,
DeFelice:2011bh,
Shirai:2012iw,Tretyakov:2012zz,Germani:2012qm,Sampurnanand:2012wy,
Zabat:2011zz},
inflation
\cite{Creminelli:2010ba,Kobayashi:2010cm,Burrage:2010cu,
Liu:2011ns,Kobayashi:2011nu,Ohashi:2012wf,Choudhury:2012yh,
Hinterbichler:2012fr} and
non-Gaussianities
\cite{Mizuno:2010ag,RenauxPetel:2011dv,Gao:2011qe,RenauxPetel:2011uk,
Choudhury:2012kw}, the
reheating of the post-inflationary universe
\cite{LevasseurPerreault:2011mw}, the growth history
\cite{Kobayashi:2010wa,Hirano:2011wj,Bellini:2012qn}, the cosmological
bounce \cite{Qiu:2011cy,Easson:2011zy,Cai:2012va}, the
cosmological perturbations
\cite{Kobayashi:2009wr,DeFelice:2010as,Gao:2011mz,Liu:2012ww,
Barreira:2012kk}, the spherical
solutions
\cite{Goon:2010xh,Babichev:2010kj,Hui:2012qt,Deser:2012gm,Rinaldi:2012vy,
Curtright:2012uz}, the
stability issues
\cite{Andrews:2010km,Endlich:2011vg,Evslin:2011rj,Appleby:2011aa,
Masoumi:2012vz,Zhou:2012fk}, and also he can use observational data
in order to constrain the parameters of the theory
\cite{Ali:2010gr,Burrage:2010rs,Nesseris:2010pc,Brax:2011sv,
DeFelice:2011aa,Hossain:2012qm,Iorio:2012pv,Appleby:2012ba,Okada:2012mn}.
Moreover, one can study and
extend the properties of Galileon theory itself
\cite{Deffayet:2010zh,Padilla:2010de,Padilla:2010ir,Padilla:2010tj,
DeFelice:2010nf,Zhou:2010di,VanAcoleyen:2011mj,
Khoury:2011da,Burrage:2011bt,Trodden:2011xh,Evslin:2011vh,
Kaloper:2011qc,Sivanesan:2011kw,Zhou:2011ix,
Curtright:2012ph,Netto:2012hm,Ali:2012cv,Babichev:2012qs,Creminelli:2012my,
Padilla:2012dx,Chu:2012kz}, and
 examine the relation of Galileons with other frameworks
\cite{deRham:2010eu,Goon:2011qf,deRham:2011by,Gannouji:2011qz,Goon:2011xf,
Burrage:2011cr,Goon:2012mu,Goon:2012dy,deRham:2012az,Charmousis:2012dw,
Gabadadze:2012tr,Zumalacarregui}.

Since Galileon cosmology exhibits interesting phenomenological features,
in the present work we perform a phase-space and stability analysis of
such a scenario, investigating in a systematic way the possible
cosmological behaviors, focusing on the late-time stable solutions. Such an
approach allows us to bypass the high non-linearities of the cosmological
equations, which prevent any complete analytical treatment, obtaining a
qualitative description of the global dynamics of these models, which is
independent of the initial conditions and the specific evolution of the
universe. Moreover, in these asymptotic solutions we calculate various
observable quantities, such as the dark energy density and
equation-of-state parameters and the deceleration parameter, and
in order to ensure that these solutions are free of ghosts and Laplacian
instabilities we additionally calculate the relevant perturbations
quantities. Interestingly enough, our analysis shows that Galileon
cosmology does not exhibit any new stable late-time solutions
apart from those of standard quintessence, and moreover the corresponding
observables are always independent of the Galileon terms. Thus,
although the Galileons can play an
important role at inflationary or at recent times, in the future, when the
universe will asymptotically reach its stable state, they will not have any
effect on its evolution.

The plan of the work is the following: In section \ref{galmodel}
we briefly review the Galileon cosmological paradigm and in section
\ref{phasespaceanalysis} we perform a detailed phase-space analysis. In
section \ref{implications} we discuss the cosmological implications and the
physical behavior of the scenario. Finally, in section \ref{Conclusions}
we summarize the obtained results.

\section{Generalized Galileon cosmology}
\label{galmodel}

In this section we briefly review Galileon cosmology for the most
generalized
scenario, presenting the background cosmological equations and the
conditions for the absence of instabilities
\cite{DeFelice:2010nf,Deffayet:2011gz,DeFelice:2011bh}. As it is known, in
order to avoid the Ostrogradski instability \cite{Ost}
it is desirable to keep the equations of motion at second order in
derivatives, and thus the most general 4-dimensional scalar-tensor theories
having second-order field equations are described by the Lagrangian
\cite{DeFelice:2011bh}
\begin{equation}
{\cal L}=\sum_{i=2}^{5}{\cal L}_{i}\,,\label{Lagsum}
\end{equation}
 where
\begin{align}
&{\cal L}_{2} = K(\phi,X),\label{eachlag2}\\
&{\cal L}_{3} = -G_{3}(\phi,X)\Box\phi,\\
&{\cal L}_{4} = G_{4}(\phi,X)\,
R+G_{4,X}\,[(\Box\phi)^{2}-(\nabla_{\mu}\nabla_{\nu}\phi)\,(\nabla^{\mu}
\nabla^{\nu}\phi)]\,,\\
&{\cal L}_{5} = G_{5}(\phi,X)\,
G_{\mu\nu}\,(\nabla^{\mu}\nabla^{\nu}\phi)\,\nonumber\\&\ \ \
\ \ \ \ -\frac{1}{6}\,
G_{5,X}\,[(\Box\phi)^{3}-3(\Box\phi)\,(\nabla_{\mu}\nabla_{\nu}\phi)\,
(\nabla^{\mu}\nabla^{\nu}\phi)+2(\nabla^{\mu}\nabla_{\alpha}\phi)\,(\nabla^
{\alpha}\nabla_{\beta}\phi)\,(\nabla^{\beta}\nabla_{\mu}\phi)]\,.\label{
eachlag5}
\end{align}
The functions $K$ and $G_{i}$ ($i=3,4,5$) depend on the scalar field $\phi$
and its kinetic energy
$X=-\partial^{\mu}\phi\partial_{\mu}\phi/2$, while
$R$ is the Ricci scalar, and $G_{\mu\nu}$ is the Einstein tensor.
$G_{i,X}$ and $G_{i,\phi}$ ($i=3,4,5$) respectively correspond to the
partial
derivatives of $G_{i}$ with respect to $X$ and $\phi$,
namely $G_{i,X}\equiv\partial G_{i}/\partial X$ and
$G_{i,\phi}\equiv\partial
G_{i}/\partial\phi$.
We mention that the above Lagrangian was first discovered by Horndeski
\cite{Horndeski} in a different but equivalent  form
\cite{Kobayashi:2011nu}.

Apart from the above scalar-tensor sectors, in a realistic cosmological
scenario one needs to take into account the matter
content of the universe, described by the Lagrangian
${\cal{L}}_m$, corresponding a perfect fluid with energy density $\rho_m$
and pressure $p_m$.
Then the total action is given by
\begin{equation}
S=\int d^{4}x\sqrt{-g}\left({\cal L}+{\cal L}_m\right)\,,\label{action1}
\end{equation}
 where $g$ is the determinant of the metric $g_{\mu\nu}$.

Let us make here an important comment on the above total action, in which
we will focus on the present work. In particular, we do not include a
possible coupling between the scalar field and the matter sector. This
is usual in many cosmological works
\cite{Silva:2009km,DeFelice:2010pv,Shirai:2012iw,Sampurnanand:2012wy,
Kobayashi:2010wa,Hirano:2011wj,Bellini:2012qn,Kobayashi:2009wr,
DeFelice:2010as,Barreira:2012kk,Nesseris:2010pc,Appleby:2012ba,
DeFelice:2011aa,Hossain:2012qm,Okada:2012mn}. However, we mention that in
its original incarnation the Galileon arises non-minimally coupled to
matter \cite{Nicolis:2008in,Deffayet:2009wt,Deffayet:2010qz}. Therefore,
strictly speaking, in the present work we view the Galileon theory as a
scalar-field, dark-energy, construction, and not as a modified gravity.
However, we have in mind that a possible coupling between the Galileon and
the matter sector, which allows for the realization of the Vainshtein
mechanism, could lead to significantly different cosmological
behavior.\footnote{We thank the referee for this comment.}

In the following we impose a flat Friedmann-Robertson-Walker (FRW)
background metric of the form
$ds^{2}=-N^{2}(t)dt^{2}+a^{2}(t)d{\bm{x}}^{2}$,
where $t$ is the cosmic time,  $x^i$ are the comoving spatial
coordinates, $N(t)$ is the lapse function, and
$a(t)$ is the scale factor. Varying   the action (\ref{action1}) with
respect to   $N(t)$ and $a(t)$ respectively, and setting $N=1$, we obtain
\begin{eqnarray}
   &&
2XK_{,X}-K+6X\dot{\phi}HG_{3,X}-2XG_{3,\phi}-6H^{2}G_{4}+24H^{2}X(G_{4,X}
+XG_{4,XX})\nonumber \\
   &&-12HX\dot{\phi}\, G_{4,\phi X}-6H\dot{\phi}\,
G_{4,\phi}
+2H^{3}X\dot{\phi}\left(5G_{5,X}+2XG_{5,XX}\right)\nonumber \\
   &&-6H^{2}X\left(3G_{5,\phi}
+2XG_{5,\phi X}\right)=
-\rho_m\,,\label{be1}
\end{eqnarray}
\begin{eqnarray}
   && K-2X(G_{3,\phi}+\ddot{\phi}\,
G_{3,X})+2(3H^{2}+2\dot{H})G_{4}-12H^{2}XG_{4,X}-4H\dot{X}G_{4,X}-8\dot{H}
XG_{4,X}\nonumber \\
  &&-8HX\dot{X}G_{4,XX}
+2(\ddot{\phi}+2H\dot{\phi})G_{4,\phi}+4XG_{4,\phi\phi}+4X(\ddot{\phi}
-2H\dot{\phi})G_{4,\phi
X}\nonumber \\
&&-2X(2H^{3}\dot{\phi}+2H\dot{H}\dot{\phi}+3H^{2}\ddot{\phi})G_{5,X}-4H^{2}
X^{2}\ddot{\phi}\, G_{5,XX} +4HX(\dot{X}-HX)G_{5,\phi
X}\nonumber \\
   &&+2[2(\dot{H}X+H\dot{X})+3H^{2}X]G_{5,\phi}+4HX\dot{\phi}\,
G_{5,\phi\phi}=-p_m\,,\label{be2}
\end{eqnarray}
 where dots denote derivatives with respect to $t$, and we also
defined the Hubble parameter $H\equiv\dot{a}/a$. Variation of
(\ref{action1}) with respect to $\phi(t)$ provides its evolution equation
\begin{equation}
\frac{1}{a^{3}}\frac{d}{dt}\left(a^{3}J\right)=P_{\phi}\,,\label{fieldeq}
\end{equation}
 with
\begin{eqnarray}
J & \equiv & \dot{\phi}K_{,X}+6HXG_{3,X}-2\dot{\phi}\,
G_{3,\phi}+6H^{2}\dot{\phi}(G_{4,X}+2XG_{4,XX})-12HXG_{4,\phi X}\nonumber
\\
 &  & +2H^{3}X(3G_{5,X}+2XG_{5,XX})-6H^{2}\dot{\phi}(G_{5,\phi}+XG_{5,\phi
X})\,,\\
P_{\phi} & \equiv & K_{,\phi}-2X\left(G_{3,\phi\phi}+\ddot{\phi}\,
G_{3,\phi X}\right)+6(2H^{2}+\dot{H})G_{4,\phi}+6H(\dot{X}+2HX)G_{4,\phi
X}\nonumber \\
 &  & -6H^{2}XG_{5,\phi\phi}+2H^{3}X\dot{\phi}\, G_{5,\phi X}\,.
\end{eqnarray}
Finally,  the evolution equation for matter takes the standard form
\begin{eqnarray}
\dot{\rho}_m+3H(\rho_m+p_m) & = & 0\,\label{rhoreq}.
\end{eqnarray}
We mention here that the four equations (\ref{be1}), (\ref{be2}),
(\ref{fieldeq}) and (\ref{rhoreq}), are not independent due to the Bianchi
identities. In particular, the scalar  equation (\ref{fieldeq}) can be
acquired
from the other three equations \cite{DeFelice:2011bh}.

In order to be able to perform the dynamical analysis of Galileon cosmology
we need to focus on some more specific models.
One  class of Galileon scenarios  has the above
Lagrangian   with the ansatzes:
\begin{eqnarray}\label{model}
K(\phi,X)=X-V(\phi),~G_3(\phi,X)=-g(\phi)X,~G_4(\phi,X)=
\frac{1}{2}\frac{1}{8\pi G}, ~G_5(\phi,X)=0,
\end{eqnarray}
corresponding to the action
\begin{equation}
S=\int d^{4}x\sqrt{-g}\left[ \frac{1}{16\pi
G}R-\frac{1}{2}\partial^{\mu}\phi\partial_{\mu}\phi-V(\phi)
-\frac{1}{2}g(\phi)\partial^{\mu}\phi\partial_{\mu}\phi\Box\phi  +{\cal
L}_m\right]\,.\label{action}
\end{equation}
Such an action is able to capture the basic, quite general, and more
interesting Galileon terms (one could straightforwardly include ansatzes
with higher powers of $X$, such as the covariant Galileon model
\cite{Deffayet:2009wt}, however for simplicity we remain to the above
simple but non-trivial Galileon action).
In this case, the gravitational field equations (\ref{be1}) and
(\ref{be2}) become
\begin{eqnarray}
\label{FR1}
&&
H^{2}=\frac{8\pi G}{3}\Big(\rho_{DE}+\rho_{m}\Big),
\\
\label{FR2}
&&
\dot{H}=-4\pi G \Big(\rho_{DE}+p_{DE}+\rho_{m}+p_{m}
\Big),
\end{eqnarray}
where we have defined the effective dark energy sector with energy density
and pressure respectively:
\begin{eqnarray}
\label{rhode}
&&\rho_{DE}=\frac{\dot{\phi}^2}{2}\left(1-6gH\dot{\phi}+g_{,\phi}\dot{\phi}
^ {
2 }\right )+V(\phi)~ ,\\
\label{pde}
&& p_{DE}=
\frac{\dot{\phi}^2}{2}
\left(1+2g\ddot{\phi}+g_{,\phi}\dot{\phi}^2
\right)-V(\phi).
 \end{eqnarray}
The scalar field equation (\ref{fieldeq}) becomes
\begin{eqnarray}
\ddot{\phi}+3H\dot{\phi}+2g_{,\phi}\dot{\phi}^{2}\ddot{
\phi}+\frac{1}{2}g_{,\phi\phi}\dot{\phi}^{4}
-3g\dot{H}\dot{\phi}^{2}-6gH\dot{\phi}\ddot{\phi}-9gH^{2}\dot{
\phi}^{2}+V_{,\phi}=0~,&
\label{fieldeq2}
\end{eqnarray}
 and we can immediately see that using (\ref{rhode}),(\ref{pde})
it can be rewritten to the standard form
\begin{eqnarray}
\dot{\rho}_{DE}+3H(\rho_{DE}+p_{DE})  =  0.\label{rhoreqde}
\end{eqnarray}
Furthermore, we can define the dark energy equation-of-state parameter
as
\begin{eqnarray}\label{EoS}
w_{DE}\equiv\frac{p_{DE}}{\rho_{DE}}=\frac{\frac{\dot{\phi}^2}{2}
\left(1+2g\ddot{\phi}+g_{,\phi}\dot{\phi}^2
\right)-V(\phi)}{\frac{\dot{\phi}^2}{2}\left(1-6gH\dot{\phi}+g_{,\phi}\dot{
\phi}
^ {
2 }\right )+V(\phi)}.
\end{eqnarray}
One can clearly see that in this scenario, according to the form of
$g(\phi)$,  $w_{DE}$ can
be quintessence-like, phantom-like, or experience the phantom divide
crossing during the evolution, which is a great advantage of Galileon
cosmology.

Without loss of generality in the following
we restrict the analysis to the dust matter case, that is we assume that
$w_m\equiv p_m/\rho_m=0$. In this case it is convenient to introduce two
additional quantities with great physical
significance, namely the ``total'' equation-of-state parameter
\begin{equation}
\label{wtot0}
w_{tot}\equiv\frac{p_{DE}}{\rho_{DE}+\rho_m}=\frac{8\pi
G}{3H^2}\left[\frac{\dot{\phi}^2}{2}
\left(1+2g\ddot{\phi}+g_{,\phi}\dot{\phi}^2
\right)-V(\phi)\right],
\end{equation}
and the deceleration parameter
\begin{equation}
q\equiv -1-\frac{\dot{H}}{H^2}=\frac{1}{2}+\frac{3}{2}w_{tot}.
\label{decc0}
\end{equation}

We close this section by mentioning that in order for the above scenario
to be free of ghosts and Laplacian instabilities, and thus cosmologically
viable, two conditions must be satisfied
\cite{DeFelice:2010pv,DeFelice:2011bh,Appleby:2011aa}. In particular, using
the
ansatzes (\ref{model}) and units where $\kappa\equiv 8\pi G=1$, these
write as \cite{DeFelice:2011bh}
\begin{equation}
c_{S}^{2}\equiv\frac{6w_{1}H-3w_{1}^{2}
-6\dot{w}_{1}-6\rho_m}{
4w_{2}+9w_{1}^{2}}\geq0\label{cscon},
\end{equation}
for the avoidance of Laplacian instabilities
associated with the scalar field propagation speed,
 and
  \begin{equation}
Q_{S}\equiv\frac{(4w_{2}+9w_{1}^{2})}{3w_{1}^{2}}>0\,,
\label{Qscon}
\end{equation}
for the absence of ghosts,
 where in our case
\begin{eqnarray}
&&w_{1} \equiv g\dot{\phi}^3+2 {H},\\
&&w_{2}  \equiv
3\dot{\phi}^2\left[\frac{1}{2}+g_{,\phi}\dot{\phi}^2-6Hg\dot{\phi}\right]
- {9H^2}.
\end{eqnarray}
Finally, we stress that according to (\ref{EoS})
and (\ref{cscon}),(\ref{Qscon}) the phantom phase can be free of
instabilities
and thus cosmologically viable, as it was already shown for Galileon
cosmology \cite{DeFelice:2011bh}.

\section{Phase space analysis}
\label{phasespaceanalysis}

In this section we perform a detailed phase-space and stability analysis
of generalized
Galileon cosmology. As usual, we first transform the
above dynamical system into its
autonomous form $\label{eomscol} \textbf{X}'=\textbf{f(X)}$
\cite{Copeland:1997et,Ferreira:1997au,Gong:2006sp,Chen:2008ft,Leon:2009rc},
where $\textbf{X}$ is the column vector constituted by suitably chosen
auxiliary variables, \textbf{f(X)} the corresponding  column
vector of the autonomous equations, with primes corresponding to
derivatives with respect to $M=\ln a$. Next we extract its critical points
$\bf{X_c}$ demanding $\bf{X}'|_{\bf{X}=\bf{X_c}}=0$, and in order to
determine their stability properties we expand around $\bf{X_c}$ as
$\bf{X}=\bf{X_c}+\bf{U}$, with $\textbf{U}$ the column vector of the
perturbations. Therefore, for each critical point we expand the
perturbation equations up to first order as $\label{perturbation}
\textbf{U}'={\bf{Q}}\cdot\textbf{U}$, where the matrix ${\bf {Q}}$ contains
the coefficients of the perturbation equations. Lastly, the eigenvalues of
${\bf {Q}}$ calculated at each critical point determine its type and
stability.

In the scenario at hand we introduce the auxiliary variables:
\begin{eqnarray}
&&x=\frac{\kappa\dot{\phi}}{\sqrt{6}H}\nonumber\\
&&y=\frac{\kappa\sqrt{V(\phi)}}{\sqrt{3}H}\nonumber\\
&&z=g(\phi) H \dot\phi\nonumber\\
&&v=\frac{1}{\phi}. \label{auxiliary}
\end{eqnarray}
Using these variables the Friedmann equation \eqref{FR1} becomes
\begin{equation}\label{Fr1norm}
(1-6 z) x^2+y^2+\frac{\sqrt{6} z g'(\phi) x^3}{g(\phi)}+\frac{ \rho_m}{3 H^2}=1.
\end{equation}
Moreover, using (\ref{Fr1norm})
and (\ref{rhode}) we can write the matter and dark energy density
parameters as:
\begin{eqnarray}
&&\Omega_m\equiv\frac{\rho_m}{3H^{2}}=1-\left[(1-6 z) x^2+y^2+\frac{\sqrt{6} z g'(\phi) x^3}{g(\phi)}\right]\nonumber\\
 &&\Omega_{DE}\equiv\frac{\kappa^{2}\rho_{DE}}{3H^{2}}=(1-6 z)
x^2+y^2+\frac{\sqrt{6} z g'(\phi) x^3}{g(\phi)}.
 \label{Omegas}
\end{eqnarray}
Note that in the limit $g(\phi)\rightarrow 0$ the above quantities are
well-defined, and they coincide with the usual quintessence ones
\cite{Copeland:1997et}.

In order to proceed forward we need to consider a specific scalar-field
potential $V(\phi)$ and a specific coupling function $g(\phi)$ with the
Galileon term. Concerning  $V(\phi)$ the usual assumption in dynamical
investigations in the literature is to assume an exponential potential of
the form
\begin{equation}
\label{exppot}
V(\phi)=V_0 e^{\lambda_V\phi},
\end{equation}
since exponential potentials are known to be significant in
various cosmological models
\cite{Schmidt:1990gb,Muller:1989rp,Copeland:1997et,Ferreira:1997au,
Gong:2006sp,Chen:2008ft,Leon:2009rc} (equivalently, but more
generally, we could use potentials satisfying
$\lambda_V=-\frac{1}{
V(\phi)}\frac{\partial V(\phi)}{\partial\phi}\approx const$, which is the
case
for arbitrary but nearly flat potentials
\cite{Scherrer:2007pu,Scherrer:2008be,Setare:2008sf}). Concerning
$g(\phi)$, and in order to remain general, we will consider two ansatzes,
namely the exponential one
 \begin{equation}
\label{expg}
g(\phi)=
g_0 e^{\lambda_g \phi},
\end{equation}
and the power-law one
 \begin{equation}
\label{powg}
 g(\phi)=g_0 \phi^n.
\end{equation}
The corresponding analysis will be performed separately in the following
two subsections.

\subsection{Scenario 1: Exponential potential and exponential coupling
  function}

In this subsection we consider the exponential potential (\ref{exppot})
and the exponential coupling function (\ref{expg}). In
this case, using the auxiliary variables (\ref{auxiliary}), the equations
\eqref{FR1}, \eqref{FR2} and \eqref{fieldeq2} can be transformed to the
autonomous form
\begin{align}
&x'=\left[18 z^2 x^3+4 \sqrt{6} z \lambda_g x^2+(2-12 z) x\right]^{-1}
\left\{18 z^2 \lambda_g^2 x^6+\left(9 \sqrt{6} z \lambda_g-27 \sqrt{6} z^2
\lambda_g\right) x^5\right. \nonumber\\
& \left. \ \ \ \ \ \ \ \ \ \ \ \ \ \ \ \ \ +\left[54
   z^2+\left(-6 \lambda_g^2-45\right) z+3\right] x^4   +\left[6 \sqrt{6} z
\lambda_g-3 \sqrt{6} y^2 z (2
   \lambda_g+\lambda_V)\right] x^3\right. \nonumber\\
 & \left. \ \ \ \ \ \ \ \ \ \ \ \ \ \ \ \
\ +\left[(27 z-3) y^2-9 z-3\right] x^2-\sqrt{6} y^2 \lambda_V
x\right\},\label{eqx}
   \end{align}
   \begin{align}
&y'= \left[18 z^2 x^3+4 \sqrt{6} z \lambda_g x^2+(2-12 z) x\right]^{-1}
\left\{18 y z^2 \lambda_g^2 x^5+y x^4
\left[9 \sqrt{6} (\lambda_V-2 \lambda_g) z^2+9 \sqrt{6} \lambda_g
   z\right] \right. \nonumber\\
& \ \ \ \ \ \ \ \ \ \ \ \ \ \ \ \ \ \ \ \ \ \     +y x^3
\left[54 z^2+12 (\lambda_g \lambda_V-3) z+3\right] +x \left[(18 z-3)
y^3+(3-18 z) y\right]
\nonumber\\
& \ \ \ \ \ \ \ \ \ \ \ \ \ \ \ \ \ \ \ \ \  \     \left.+ x^2
\left\{y \left[6 \sqrt{6} z
   (\lambda_g-\lambda_V)+\sqrt{6} \lambda_V\right]-3 \sqrt{6} y^3 z (2
 \lambda_g+\lambda_V)\right\} \right\}, \label{eqy}
   \end{align}
   \begin{align}
& z'= \left[18 z^2 x^3+4 \sqrt{6} z \lambda_g x^2+(2-12 z) x\right]^{-1}
\left\{-18 z^3 \lambda_g^2 x^5+\left(27 \sqrt{6} z^3 \lambda_g-9 \sqrt{6}
z^2 \lambda_g\right) x^4\right. \nonumber\\
 &  \ \ \ \ \ \ \   \ \ \ \ \  \ \ \ \ \,  +x^3\left[-54
   z^3+9 \left(2 \lambda_g^2+3\right) z^2-3 z\right]
+ x \left[\left(3 z-9 z^2\right) y^2+27
   z^2-9 z\right]  \nonumber\\
 & \left. \ \ \ \ \ \ \   \ \ \ \ \  \ \ \, \ \  +x^2\left[-18 \sqrt{6}
\lambda_g z^2+3 \sqrt{6} y^2 (2
   \lambda_g+\lambda_V) z^2+2 \sqrt{6} \lambda_g z\right]    -\sqrt{6} y^2
z \lambda_V \right\} \label{eqz},
\end{align}
defined in the (non-compact) phase space $\Psi=\left\{(x,y,z): y>0, z
\left(\sqrt{6} x \lambda_g-6\right) x^2+x^2+y^2\leq 1\right\}.$
Note that in this case the variable $v$ is not needed.

Using (\ref{rhode})  and the Friedmann equation (\ref{Fr1norm}), we can
write the density parameters as:
\begin{eqnarray}
&&\Omega_m\equiv\frac{\rho_m}{3H^{2}}=1-\left[(1-6 z) x^2+y^2+{\sqrt{6} z x^3}\lambda_g\right]\nonumber\\
 &&\Omega_{DE}\equiv\frac{\kappa^{2}\rho_{DE}}{3H^{2}}=(1-6 z)
x^2+y^2+{\sqrt{6} z x^3}\lambda_g,
 \label{OmegasExp}
\end{eqnarray}
while for the dark-energy equation-of-state parameter (\ref{EoS}) we
obtain:
{\small \begin{equation}\label{wdephase}
w_{DE}=\frac{6 z^2 \lambda_g^2 x^4+3 \sqrt{6} (1-2 z) z \lambda_g x^3+[3 z
(3 z-4)+1] x^2-\sqrt{6} y^2 z (2
   \lambda_g+\lambda_V) x+y^2 (6 z-1)}{\left[zx^2 \left(\sqrt{6} x
\lambda_g-6\right) +x^2+y^2\right]
   \left[z \left(9 z x^2+2 \sqrt{6} \lambda_g x-6\right)+1\right]}.
\end{equation}}
As we mentioned above, according to the variable $z$, that is according to
the coupling function $g(\phi)$, in this scenario $w_{DE}$ can
be quintessence-like, phantom-like, or experience the phantom divide
crossing during the evolution.
Furthermore, the total equation-of-state parameter (\ref{wtot0})
becomes
{\small
\begin{align}\label{wtot}
&w_{tot}\equiv\frac{p_{DE}}{\rho_{DE}+\rho_m}=w_{DE}\Omega_{DE}=
\nonumber\\&
=\frac{6 z^2 \lambda_g^2 x^4+3 \sqrt{6} (1-2 z) z \lambda_g x^3+[3 z (3
z-4)+1] x^2-\sqrt{6} y^2 z (2
   \lambda_g+\lambda_V) x+y^2 (6 z-1)}{z \left(9 z x^2+2 \sqrt{6} \lambda_g x-6\right)+1},
\end{align}}
and the deceleration parameter (\ref{decc0}) reads
 \begin{eqnarray}
q&\equiv& -1-\frac{\dot{H}}{H^2}=\frac{1}{2}+\frac{3}{2}w_{tot}\nonumber\\
&=& \left\{2 z \left(9 z x^2+2 \sqrt{6}
   \lambda_g x-6\right)+2\right\}^{-1}
\left\{9 \sqrt{6} (1-2 z) z \lambda_g x^3+[36 (z-1)
z+3] x^2\right.\nonumber\\
&&\left.\ \ \ \ \ \ \ \ \ \  +18 z^2 \lambda_g^2 x^4+\sqrt{6} z
\left[\left(2-6
   y^2\right) \lambda_g-3 y^2 \lambda_V\right] x+\left(3 y^2-1\right) (6
z-1)\right\}.
\label{decc}
\end{eqnarray}
Finally, the two instability-related quantities (\ref{cscon})
and (\ref{Qscon}) are respectively written as
 \begin{eqnarray}
\label{csmod1}
&c_S^2= \left\{{x \left[z
   \left(9 z x^2+2 \sqrt{6} \lambda_g x-6\right)+1\right]^2}\right\}^{-1}
\left\{3 \sqrt{6} z^3 \lambda_g x^4+3 z^2x^3\left(2
\lambda_g^2-6 z+5\right) \right.\nonumber\\
&\left.\ \ \ \ \ \  -27 z^4 x^5+2 \sqrt{6} (1-4 z) z
   \lambda_g x^2+\left\{z \left[3 z\left(5-3 y^2\right) -4\right]+1\right\}
x+\sqrt{6} y^2 z \lambda_V\right\}
\end{eqnarray}
 and
 \begin{align}
\label{Qsmod1}
 Q_S=\frac{3 x^2 \left[z \left(9 z x^2+2 \sqrt{6} \lambda_g
x-6\right)+1\right]}{\left(3 z x^2+1\right)^2}.
 \end{align}

\subsubsection{Finite phase-space analysis}\label{model1finite}

Let us now proceed to the phase-space analysis. The real and physically
meaningful critical points $(x_c, y_c, z_c)$ of the autonomous system
(\ref{eqx})-(\ref{eqz}) (that is corresponding to an expanding universe,
and  thus  possessing $H>0$, with $0\leq\Omega_{DE}\leq1$), are obtained by
setting the left hand sides of the
equations  to zero, and they are presented in Table \ref{crit}. In the same
table
we provide their existence conditions.
For each critical point of Table \ref{crit} we calculate the $3\times3$
matrix ${\bf {Q}}$ of the linearized perturbation equations of the system
(\ref{eqx})-(\ref{eqz}), and in order to determine the type and stability
of the point we examine the sign of the real part of the eigenvalues of
${\bf {Q}}$. The details of the analysis and the various eigenvalues are
presented in Appendix \ref{appmod1a}, and in Table \ref{crit} we
summarize the stability results. Moreover,
for each critical point we calculate the values of $\Omega_{DE}$,
$w_{DE}$, $w_{tot}$ and $q$  given by (\ref{OmegasExp})-(\ref{decc}),
as well as the instability-related quantities $c_S^2$ and $Q_S$ given in
(\ref{csmod1}),(\ref{Qsmod1}), and we summarize the results in  Table
\ref{crit1}.
\begin{table*}[!]
\begin{center}
\begin{tabular}{|c|c|c|c|c|c|}
\hline
 {\small{Cr. P.}}& $x_c$ & $y_c$ &  $z_c$ & Exist for & Stability
\\
\hline \hline
$A^+$& $+1$ & $0$ & $0$& always & {\small{unstable for
$\lambda_V>-\sqrt{6}$,$\lambda_g>\sqrt{6}$ }}
\\
&&&& & saddle point otherwise \\
\hline
$A^-$& $-1$ & $0$ & $0$& always & {\small{unstable for
$\lambda_V<\sqrt{6}$,$\lambda_g<-\sqrt{6}$ }}
\\
&&&& & saddle point otherwise \\
\hline
&&&& &  {\small{unstable for
$\lambda_g>\sqrt{6}$,$\lambda_V>-\lambda_g$ }} \\
 $B^+$& {\small{$\frac{ \lambda_{{g}}+\sqrt
{{\lambda_{{g}}}^{2}-6}}{\sqrt{6}}$}} & $0$ & {\small{ $\frac{
3-\alpha^-(\lambda_g) }{9}$}} & $\lambda_g^2\geq 6$    & saddle
point otherwise  \\
\hline
&&&& &  {\small{unstable for
$\lambda_g<-\sqrt{6}$,$\lambda_V<-\lambda_g$ }}   \\
 $B^-$& {\small{$\frac{\lambda_{{g}}-\sqrt
{{\lambda_{{g}}}^{2}-6}}{\sqrt {6}}$}} & $0$ &{\small{ $\frac{
3-\alpha^+(\lambda_g)}{9}$}} & $\lambda_g^2\geq 6$    & saddle
point otherwise
  \\
\hline
&&&& & {\small{stable node for
$-\sqrt{3}<\lambda_V<0$,$\lambda_g<-\lambda_V$ }}   \\
 $C$& $-\frac{ \lambda_{{V}}}{\sqrt {6}}$ & {\small{$\sqrt
{1-\frac{{\lambda_{{V}}}^{2}}{6}}$ }} & $0$ &{\small{ $0<\lambda_V^2\leq
6$}} &
{\small{stable node for
$0<\lambda_V<\sqrt{3}$,$\lambda_g>-\lambda_V$ }}
 \\
&&&& &  saddle
point otherwise\\
\hline
 $C_0$& $0$ & $1$  & $0$ & $\lambda_V=0$ &
stable node \\
\hline
&&&& & {\small{stable node for
$-\sqrt{\frac{24}{7}}\leq\lambda_V<-\sqrt{3}$,$\lambda_g<-\lambda_V$ }}  \\
$D$& $-{\frac {\sqrt {6}}{2\lambda_{{V}}}}$ & ${\frac {\sqrt
{6}}{2\lambda_{{V}}}}$ & 0 & $\lambda_V^2\geq 3$ &  {\small{stable node for
$\sqrt{3}<\lambda_V<\sqrt{\frac{24}{7}}$,$\lambda_g>-\lambda_V$ }} \\
&&&& & {\small{stable spiral for
$\lambda_V<-\sqrt{\frac{24}{7}}$,$\lambda_g<-\lambda_V$ }}  \\
&&&& & {\small{stable spiral for
$\lambda_V>\sqrt{\frac{24}{7}}$,$\lambda_g>-\lambda_V$ }}  \\ \hline
$O_1$ &0&0&0& always & saddle \\
\hline
\end{tabular}
\end{center}
\caption[crit]{\label{crit}Scenario 1: Exponential potential and
exponential coupling
  function. The real and physically meaningful
critical points of the autonomous system \eqref{eqx}-\eqref{eqz},
and their existence and stability
conditions. We have introduced the functions
$\alpha^\pm(\lambda_g)=\lambda_g^2\pm\lambda_g\sqrt{\lambda_g^2-6}.$}
\end{table*}
\begin{table*}[ht]
\begin{center}
\begin{tabular}{|c|c|c|c|c|c|c|}
\hline
 Cr. P. & $\Omega_{DE}$ &  $w_{DE}$ & $w_{tot}$ & q & $c_S^2$ &
$Q_S$  \\
\hline \hline
$A^\pm$ & $1$ & $1$& $1$& $2$& $1$& $3$
\\
\hline
&&&& &&  \\
 $B^\pm$&  $1$ & $-1 +\frac{\alpha^\pm(\lambda_g)}{3}$&
$-1 +\frac{\alpha^\pm(\lambda_g)}{3}$& $-1
+\frac{\alpha^\pm(\lambda_g)}{2}$ & $0$& arbitrary  \\
\hline
&&&& &&   \\
 $C$&  $1$  & $-1 +\frac{\lambda_V^2}{3}$
& $-1 +\frac{\lambda_V^2}{3}$ & $-1 +\frac{\lambda_V^2}{2}$ & $1$
&$\frac{\lambda_V^2}{2}$\\
\hline
 $C_0$&  $1$  & $-1$
& $-1$ & $-1 $ & $1$
&$0$\\
\hline
&&&& &&   \\
$D$&  $\frac{3}{\lambda_V^2}$ & 0 & 0 &$\frac{1}{2}$ &
$1$ & $\frac{9}{2 \lambda_V^2}$
\\
\hline
&&&& &&  \\fz
$O_1$&  0 & arbitrary & 0 &$\frac{1}{2}$ &
$1$ & $0$\\\hline
\end{tabular}
\end{center}
\caption[crit1]{\label{crit1}Scenario 1: Exponential potential and
exponential coupling
  function. The real and physically meaningful
critical points of the autonomous system \eqref{eqx}-\eqref{eqz}, and the
corresponding values of the dark-energy
density parameter $\Omega_{DE}$, of the  dark-energy equation-of-state
parameter $w_{DE}$, of the total equation-of-state parameter
  $w_{tot}$ and of the deceleration parameter $q$. The last two columns
contain the instability-related parameters $c_S^2$ and $Q_S$ (from
(\ref{csmod1}),(\ref{Qsmod1})), which must respectively be non-negative and
positive for a scenario free of ghosts and instabilities.
  We have introduced the functions
$\alpha^\pm(\lambda_g)=\lambda_g^2\pm\lambda_g\sqrt{\lambda_g^2-6}>0.$}
\end{table*}

\subsubsection{Phase-space analysis at infinity}\label{inftA}

Since the dynamical system (\ref{eqx})-(\ref{eqz})  is
non-compact, there could be
features in the asymptotic regime which are non-trivial for the
global dynamics. Therefore, in order to complete the phase-space analysis
we have to extend the investigation using the Poincar\'e
central projection method
\cite{PoincareProj}.

We consider the Poincar\'e variables
\begin{equation}
\label{Transf}
x_r=\rho \cos\theta \sin \psi ,\ z_r=\rho
\sin \theta \sin \psi ,\, y_r= \rho \cos \psi,\,
\end{equation} with $\rho=\frac{r}{\sqrt{1+r^2}},$ $r=\sqrt{x^2+y^2+z^2},$
$\theta\in[0,2\pi],$
and $-\frac{\pi}{2}\leq \psi \leq \frac{\pi}{2}$ (we
restrict the  $\psi$ angle to this range since the physical
region is given by $y>0$)
\cite{PoincareProj,Leon:2008de,Xu:2012jf,Leon2011}.   Therefore, the
points at ``infinity''
($r\rightarrow+\infty$) are those with $\rho\rightarrow 1.$
Moreover, the physical phase space is now given by
$\left(x_r,y_r,z_r\right)\in [-1,1]\times [0,1]\times[-1,1]$, such that
\begin{equation}
\frac{x_r^2+y_r^2}{1-x_r^2-y_r^2-z_r^2}-\frac{x_r^2 z_r \left(\sqrt{6}\lambda_g x_r-6
   x_r^2-6 y_r^2-6 z_r^2+6\right)}{\left(1-x_r^2-y_r^2-z_r^2\right)^{5/2}}\leq 1,
\end{equation}
and $x_r^2+y_r^2+z_r^2\leq 1.$

Inverting relations (\ref{Transf}) and substituting into
(\ref{OmegasExp}),(\ref{wdephase}), we obtain the dark energy density and
equation-of-state parameters as a function of the  Poincar\'e variables,
namely:
\begin{equation}
\Omega_{DE}=\frac{x_r^2+y_r^2}{1-x_r^2-y_r^2-z_r^2}-\frac{x_r^2 z_r
\left(\sqrt{6}\lambda_g x_r-6
   x_r^2-6 y_r^2-6
z_r^2+6\right)}{\left(1-x_r^2-y_r^2-z_r^2\right)^{5/2}},\ \ \ \ \ \ \ \ \ \
\ \ \ \ \ \ \ \ \ \ \ \
 \label{Omegas22}
\end{equation}
\begin{eqnarray}
&&w_{DE}=  \left\{\left[\sqrt{6} z_r \lambda_g
x_r^3+\left(\zeta -6 z_r
\sqrt{\zeta }\right) x_r^2+y_r^2
   \zeta \right] \left(9 x_r^2 z_r^2-6 \zeta ^{3/2} z_r+2 \sqrt{6} x_r
\zeta  \lambda_g z_r+\zeta
   ^2\right)\right\}^{-1}\nonumber\\
&&\ \ \ \ \ \ \ \ \ \ \ \ \ \ \ \ \  \zeta \left\{6 z_r^2
\lambda_g^2 x_r^4+3 \sqrt{6} z_r \left(\sqrt{\zeta }-2 z_r\right)
\sqrt{\zeta }
   \lambda_g x_r^3+\zeta  \left(9 z_r^2-12 \sqrt{\zeta } z_r+\zeta
\right) x_r^2\right.\nonumber\\
&&\left.\ \ \ \ \ \ \ \ \ \ \ \ \ \ \ \ \ \ \ \ \ \ \ \ \ \ \ \ \ \ \ \ \ \
\ \ \ \ \ \ \ \ \ \ \ -\sqrt{6} y_r^2 z_r
   \zeta  (2 \lambda_g+\lambda_V) x_r+y_r^2 \left(6 z_r-\sqrt{\zeta
}\right) \zeta
   ^{3/2}\right\},
\label{wdephase22}
\end{eqnarray}
with $\zeta=1-x_r^2-y_r^2-z_r^2$,
and similarly substituting into
(\ref{wtot}), (\ref{decc})  we obtain the corresponding expressions for
the total equation-of-state and deceleration parameters:
\begin{eqnarray}
&& w_{tot}
=\left[{\zeta  \left(9 x_r^2
   z_r^2-6 \zeta ^{3/2} z_r+2 \sqrt{6} x_r \zeta  \lambda_g z_r+\zeta
^2\right)}\right]^{-1}   \left\{3
\sqrt{6} z_r \left(\sqrt{\zeta }-2 z_r\right) \sqrt{\zeta } \lambda_g
   x_r^3 \right.
\nonumber\\
 && \ \ \ \ \ \ \ \ \ \ \ \ \ \  \ \  \ \ \ \ \ \ \ \  \ \  +\zeta
\left(9 z_r^2-12
\sqrt{\zeta } z_r+\zeta \right) x_r^2 -\sqrt{6} y_r^2 z_r \zeta  (2
   \lambda_g+\lambda_V) x_r\nonumber\\
&&\left.  \ \ \ \ \ \ \ \ \ \ \ \ \  \ \  \ \ \ \ \ \ \ \  \ \  \ \ \ \ \ \
\ \ \ \ \ \ \  \ \  \ \ \ \ \ \ \ \  \ \ \ +6 z_r^2 \lambda_g^2 x_r^4+y_r^2
\left(6 z_r-\sqrt{\zeta }\right) \zeta
^{3/2}\right\},
\label{wtotphase22}
\end{eqnarray}
\begin{eqnarray}
&&q=\left[{2 \zeta  \left(9 x_r^2 z_r^2-6 \zeta ^{3/2} z_r+2 \sqrt{6} x_r
\zeta
   \lambda_g z_r+\zeta ^2\right)}\right]^{-1}\left\{9 \sqrt{6} z_r
\left(\sqrt{\zeta }-2 z_r\right) \sqrt{\zeta }
\lambda_g x_r^3\right. \nonumber\\
 && \ \ \ \ \ \ \ \ \ \ \ \ \ \  \ \ +3 \zeta  \left(12 z_r^2-12
\sqrt{\zeta } z_r+\zeta \right) x_r^2+\sqrt{6} z_r \zeta  \left[2 \zeta
   \lambda_g-3 y_r^2 (2 \lambda_g+\lambda_V)\right] x_r \nonumber\\
&& \ \ \ \ \ \ \ \  \ \ \ \ \ \ \ \ \ \ \ \ \  \ \ \ \ \ \ \ \ \ \ \ \ \ \
\ \ \ \ \ \ \ \left.+18 z_r^2 \lambda_g^2
x_r^4+\zeta
   ^{3/2}\left(\sqrt{\zeta }-6 z_r\right)  \left(\zeta -3
y_r^2\right)\right\}.
\label{wqphase22}
\end{eqnarray}
Finally, substitution into (\ref{csmod1}),(\ref{Qsmod1})
provides the corresponding expressions for the instability-related
quantities:
\begin{eqnarray}
&&c_S^2=\left[{x_r \left(9 x_r^2 z_r^2-6 \zeta
   ^{3/2} z_r+2 \sqrt{6} x_r \zeta  \lambda_g z_r+\zeta
^2\right)^2}\right]^{-1}\left\{-27 z_r^4 x_r^5+3 \sqrt{6} z_r^3 \zeta
\lambda_g x_r^4\right. \nonumber\\
&&  \ \ \ \ \ \ \ \ \ \ \ \ \  +3 z_r^2 \zeta ^{3/2}
\left[\sqrt{\zeta }
   \left(2 \lambda_g^2+5\right)-6 z_r\right] x_r^3+2 \sqrt{6} z_r
\left(\sqrt{\zeta }-4 z_r\right) \zeta ^{5/2}
   \lambda_g x_r^2 \nonumber\\
&&  \ \ \ \ \ \ \ \ \  \ \ \ \ \ \ \ \ \ \ \ \ \  \ \ \left.+\zeta ^2
\left[\zeta
\left(15 z_r^2-4 \sqrt{\zeta } z_r+\zeta \right)-9 y_r^2
   z_r^2\right] x_r+\sqrt{6} y_r^2 z_r \zeta ^3 \lambda_V\right\},
\label{cspoinc1}
\end{eqnarray}
\begin{eqnarray}
Q_S=\frac{3 x_r^2 \left(9 x_r^2 z_r^2-6 \zeta ^{3/2} z_r+2 \sqrt{6} x_r
\zeta  \lambda_g z_r+\zeta
   ^2\right)}{\left(3 z_r x_r^2+\zeta ^{3/2}\right)^2}.
\label{Qspoinc1}
\end{eqnarray}

Therefore, performing the analysis described in Appendix \ref{Poincare1} we
conclude that there are two physical critical points at infinity,
namely $K^\pm$. These critical points,
along with their stability conditions and the corresponding
values of observables, are presented
in Table \ref{critmod1inf}.

\begin{table*}[!]
\begin{center}
\begin{tabular}{|c|c|c|c|c|c|c|c|c|c|c|}
\hline
 Cr. P. & $x_r$ & $y_r$ & $z_r$ &Stability&  $\Omega_{DE}$ &  $w_{DE}$ &
$w_{tot}$ & q & $c_S^2$ &
$Q_S$  \\
\hline \hline
 $K^\pm$&  $0$ & $0$ & $\pm 1$& {\small{unstable}} &$0$ &
{\small{arbitrary}} & $0$ &
$\frac{1}{2}$ & {\small{arbitrary}} & $0$ \\
\hline
\end{tabular}
\end{center}
\caption[crit1]{\label{critmod1inf} Scenario 1: Exponential potential and
exponential coupling
  function. The real and physically meaningful
critical points at infinity of the autonomous system
\eqref{eqx}-\eqref{eqz}, and the
corresponding values of the dark-energy
density parameter $\Omega_{DE}$, of the  dark-energy equation-of-state
parameter $w_{DE}$, of the total equation-of-state parameter
  $w_{tot}$ and of the deceleration parameter $q$. The last two columns
contain the instability-related parameters $c_S^2$ and $Q_S$ (from
(\ref{cspoinc1}),(\ref{Qspoinc1})), which must respectively be non-negative
and
positive  for a scenario free of ghosts and instabilities.
}
\end{table*}

\subsection{Scenario 2: Exponential potential and power-law coupling
  function}

Let us consider an exponential potential $V(\phi)=V_0 e^{\lambda_V \phi}$
and a power-law coupling function $g(\phi)=g_0 \phi^n$. In this case, using
the
auxiliary variables (\ref{auxiliary}),  the equations \eqref{FR1},
\eqref{FR2} and \eqref{fieldeq2} can be transformed to the autonomous form:
\begin{eqnarray}\label{eqpowerx}
&&x'=\left[2\sqrt {6}+24xvzn-12z\sqrt {6}+18\sqrt
{6}x^2z^2\right]^{-1}\left\{18\sqrt {6}n \left( n+1
\right) {v}^{2}{z}^{2}{x}^{5} -6{y}^{2}\lambda_V\right. \nonumber\\
 && \left.
\ \ \ \ \ \ \ \ \ \ \ \  + 54vznx^4
 \left(1 -3z \right) + 3\sqrt {6}x^3\left\{ 18{
z}^{2}+ \left[  2n (1-n) {v}^{2}-15
  \right] z+1 \right\} \right. \nonumber\\
&&
\left. \ \ \ \  \ \ \ \ \ \ \ \  - 18z x^2\left[  \left( 2
vn+\lambda_V \right) y^2-2vn \right]+ 3\sqrt {6}x\left[
 \left( 9z-1 \right) y^2-3z-1
 \right] \right\}
\end{eqnarray}
\begin{eqnarray}\label{eqpowery}
&& y'=\left[2\sqrt {6}+24xvzn-12z\sqrt {6}+18\sqrt
{6}{x}^{2}{z}^{2}\right]^{-1}\left\{18\sqrt {6}n \left( n+1
\right)
v^2z^2yx^4\right.  \ \ \  \ \ \ \ \ \ \ \  \nonumber\\
&&    \ \ \ \ \ \ \  \ \ \ \ \ \ \  +  3\sqrt {6}yx^2
\left[ 18 z^2+4 \left(\lambda_V nv-
3 \right) z+1  \right]+3 \sqrt {6}\left( 6z-1 \right) y^3
\nonumber\\
&& \left.  \ \ \ \ \ \ \  \ \ \ \ \ \ \  - 6 x\left\{ 3 zy^3 \left(
2 nv+ \lambda_V \right)-  y\left[  6z \left(   nv-
\lambda_V \right)+ \lambda_V \right] \right\}\right. \nonumber\\
&& \left.  \ \ \ \ \ \ \   \ \ \ \ \ \ \
 + 3 \sqrt {6} y \left( 1-6 z
 \right)+ 54yx^3 \left[  \left(\lambda_V -2 nv \right) z^2+ vzn
\right]
\right\}\end{eqnarray}
\begin{eqnarray}\label{eqpowerz}
&&z'=\left[2x \left( \sqrt {6}+12xvzn-6z\sqrt {6}+9\sqrt {6}{x}^{2}{z}^{2
} \right)
\right]^{-1}
\left\{-18\sqrt {6}n \left( n+1
\right) v^2z^3x^5\right. \ \ \  \ \ \ \ \ \
 \nonumber\\
&&    \ \ \ \ \ \ \ \ \  +
  54nvz^2 x^4\left( 3z-1\right)   -3\sqrt {6}z x^3\left\{ 18
z^2- \left[ 2n \left( 1+3n \right) {
v}^{2}+9 \right] z+1 \right\}
\nonumber\\
&& \left.  \ \ \ \ \ \ \ \ \ +
6z
x^2\left[  \left( 6nv+3\lambda_V \right) zy^2-18nvz+2vn \right]\right.
 \nonumber\\
&& \left.  \ \ \ \ \ \ \ \ \ + 3z
\sqrt {6} x \left[ y^2 \left( 1-3z \right)+9z-3
\right]
-6zy^2\lambda_V
\right\}
\end{eqnarray}
\begin{eqnarray}
&v'=-\sqrt {6}v^2x, \ \ \  \ \ \ \ \ \ \ \   \ \ \  \ \ \ \ \ \ \ \ \ \ \ \
\ \ \ \  \ \ \  \ \ \ \ \ \ \ \   \ \ \  \ \ \ \ \ \ \ \   \ \ \  \ \ \ \ \
\ \ \  \ \ \  \ \ \ \ \ \ \ \   \ \ \  \ \ \ \ \ \ \ \  \label{eqpowerw}
\end{eqnarray}
defined in the non-compact phase space   $$\Psi=\left\{(x,y,z,v): y>0, z
\left(\sqrt{6}n v x -6\right) x^2+x^2+y^2\leq 1, v\in\mathbb{R}\right\}.$$
Note that contrary to the previous Scenario 1, in the case at hand we do
need the fourth auxiliary variable $v$.

Using (\ref{Fr1norm})
and (\ref{rhode}) we can write the density parameters as:
\begin{eqnarray}
&&\Omega_m\equiv\frac{\rho_m}{3H^{2}}=1-\left[(1-6 z) x^2+y^2+{\sqrt{6}nv z
x^3}\right]\nonumber\\
 &&\Omega_{DE}\equiv\frac{\kappa^{2}\rho_{\phi}}{3H^{2}}=(1-6 z)
x^2+y^2+{\sqrt{6 }n v z x^3},
 \label{Omegaspowerlaw}
\end{eqnarray}
while for the dark-energy equation-of-state parameter (\ref{EoS}) we
obtain:
 \begin{eqnarray}\label{wdephasepowerlaw}
&&w_{DE}=\left\{\left( \sqrt {6}{x}^{2}-6 \sqrt {6}z{x}^{2}+6
v{x}^{3}zn+\sqrt {6}{y}^{2}
\right)  \left( \sqrt {6}+12 xvzn-6 z\sqrt {6}+9 \sqrt {6}{x
}^{2}{z}^{2} \right)\right\}^{-1}\nonumber\\
&&  \ \ \ \ \ \ \ \ \  \ \ \ \ \ \ \ \ \  6\left[6\left(1+n
\right)n {v}^{2}{z}^{2}{x}^{4}+ 3\sqrt{6}n\left( 1-2z\right) v{x}^{3}z+
\left( 1+9 {z}^{
2}-12 z \right) {x}^{2}\right.\nonumber\\
&&  \ \ \ \ \ \ \ \ \  \ \ \ \ \ \ \ \  \ \ \ \ \  \ \ \ \ \ \ \ \ \ \ \ \
\  \ \ \ \ \ \ \ \ \ \ \ \ \  \ \ \ \ \   \ \left. -\sqrt{6}\left( 2 n
v+6 \lambda_V \right)
z{y}^{2}x+ \left( 6 z-1 \right) {y}^{2}\right].
\end{eqnarray}
Furthermore, the total equation-of-state parameter (\ref{wtot0}) parameter
reads
\begin{eqnarray}\label{wtotpowerlaw}
&&w_{tot}= \left( 1+2 \sqrt{6}xvzn-6 z+9{x
}^{2}{z}^{2} \right)^{-1}\left[6\left(1+n \right)n {v}^{2}{z}^{2}{x}^{4}+
3\sqrt{6}n\left( 1-2z\right)v{x}^{3}z\right.\nonumber\\
&&  \ \ \ \ \  \ \ \ \ \ \ \ \ \ \ \ \
\  \ \ \ \ \  \left.+ \left( 1+9 {z}^{
2}-12 z \right) {x}^{2}- \sqrt{6}\left( 2 n v+6 \lambda_V \right)
z{y}^{2}x+ \left( 6 z-1 \right) {y}^{2}
\right],
\end{eqnarray}
and the deceleration parameter (\ref{decc0}) becomes
 \begin{equation}\label{deccpowerlaw}
q=\frac{1}{2}+\frac{3}{2}w_{tot}.
\end{equation}
Finally, from (\ref{cscon}),(\ref{Qscon}) we find
\begin{align}\label{cs2powerlaw}
&&c_S^2=\left[x \left( 1+2\sqrt{6}\,zxnv-6z+9x^2z^2
 \right) ^2
\right]^{-1}\left\{-27{x}^{5}{z}^{4}+3\sqrt {6}{x}^{4}{z}^{3}nv\right.
 \nonumber\\
 && \left.  \ \ \ \ \ \ \ + 3z^2x^3\left[ -6z+   5+ 2
n\left( n-1 \right) v^2  \right] + 2\sqrt {6} z nv\left( 1  -4 z
 \right) x^2\right. \nonumber\\
&& \left.  \ \ \ \ \ \ \ +  x\left( 15
{z}^{2}-9 {y}^{2}{z}^{2}-4 z+1
 \right)+ z\sqrt {6}{y}^{2}\lambda_V
\right\}
\end{align}
and
\begin{align}\label{QSpowerlaw}
Q_S={\frac { \left( 3+6\sqrt {6} zxnv-18z+27x^2z^2 \right)
x^2}{ \left( 3 x^2z+1
 \right)^2}}.
\end{align}

\subsubsection{Finite phase-space analysis}

Let us now proceed to the phase-space analysis. The real and physically
meaningful critical points $(x_c, y_c, z_c, v_c)$ of the autonomous system
\eqref{eqpowerx}-\eqref{eqpowerw} (that is corresponding to an expanding
universe, and  thus  possessing $H>0$, with $0\leq\Omega_{DE}\leq1$),  are
presented in Table \ref{crit2}, along with their existence conditions.
The details of the analysis and the various eigenvalues are
presented in Appendix \ref{appmod2a}. Furthermore, in Table \ref{crit22}
we display the corresponding values of the observables $\Omega_{DE}$,
$w_{DE}$, $w_{tot}$ and $q$ given by
(\ref{Omegaspowerlaw})-(\ref{deccpowerlaw}),
as well as the instability-related quantities $c_S^2$ and $Q_S$ given in
(\ref{cs2powerlaw}),(\ref{QSpowerlaw}).
\begin{table*}[ht]
\begin{center}
\begin{tabular}{|c|c|c|c|c|c|c|}
\hline
 {\small{Cr. P.}}& $x_c$ & $y_c$ &  $z_c$& $v_c$ & Exist for & Stability
\\
\hline \hline
$E^{\pm}$& $\pm1$& $0$ & $0$ & $0$& always &  saddle point
\\
\hline
&&&& & &   stable node for
$\lambda_V^2<3$   \\
$F$& $-\frac{\lambda_V}{6}$ & $\sqrt{1-\frac{\lambda_V^2}{6}}$ & $0$ &
$0$    & {\small{ $0<\lambda_V^2\leq
6$}} &
 saddle point for  $3<\lambda_V^2<6$
 \\
\hline
$F_0$& $0$ & $1$ & $0$ &
$0$    & $\lambda_V=0$ &
 stable (not asymptotically for $n\neq0$)  \\
&&&& & &
 stable (asymptotically for $n=0$)  \\
\hline
&&&& & & stable node for
$3<\lambda_V^2 <\frac{24}{7}$        \\
 $G$& $-\frac{ \sqrt{6}}{2\lambda_V}$& $\frac{ \sqrt{6}}{2\lambda_V}$ &
$0$ & $0$ &$\lambda_V^2>3$  &
 stable spiral for
$\frac{24}{7}<\lambda_V^2$  \\
\hline
$O_2$&  0 & 0 & 0 &0 & always &
saddle\\\hline
\end{tabular}
\end{center}
\caption[crit]{\label{crit2} Scenario 2: Exponential potential and
power-law coupling
  function.  The real and physically meaningful
critical points of the autonomous system \eqref{eqpowerx}-\eqref{eqpowerw},
and their existence and stability
conditions. }
\end{table*}
\begin{table*}[ht]
\begin{center}
\begin{tabular}{|c|c|c|c|c|c|c|}
\hline
&&&& &&   \\
 Cr. P.&   $\Omega_{DE}$ &  $w_{DE}$ & $w_{tot}$ & q & $c_S^2$ & $Q_S$  \\
\hline \hline
&&&& &&  \\
$E^\pm$& $1$ & $1$& $1$& $2$& $1$& $3$
\\
\hline
&&&& &&  \\
 $F$&  $1$  & $-1 +\frac{\lambda_V^2}{3}$
&$-1 +\frac{\lambda_V^2}{3}$& $-1 +\frac{\lambda_V^2}{2}$ & $1$
&$\frac{\lambda_V^2}{2}$\\
\hline
$F_0$&  $1$  & $-1$
& $-1$ & $-1$ & $1$ & 0\\
\hline
&&&& &&  \\
 $G$& $\frac{3}{\lambda_V^2}$ & 0 & 0 &$\frac{1}{2}$ &
$1$ & $\frac{9}{2 \lambda_V^2}$\\
\hline
&&&& &&  \\
$O_2$&  0 & arbitrary & 0 &$\frac{1}{2}$ &
$1$ & $0$\\\hline
\end{tabular}
\end{center}
\caption[crit]{\label{crit22} Scenario 2: Exponential potential and
power-law coupling
  function. The real and physically meaningful critical
points of the autonomous system
\eqref{eqpowerx}-\eqref{eqpowerw}, and the
corresponding values of the dark-energy
density parameter $\Omega_{DE}$, of the  dark-energy equation-of-state
parameter $w_{DE}$, of the total equation-of-state parameter
  $w_{tot}$ and of the deceleration parameter $q$. The last two columns
contain the instability-related parameters $c_S^2$ and $Q_S$ (from
(\ref{cs2powerlaw}),(\ref{QSpowerlaw})), which must respectively be
non-negative
and
positive  for a scenario free of ghosts and instabilities.
 }
\end{table*}
We mention that the stability of the above points does not
depend on the exponent $n$, since these points have $z_c=0$ in which case
$n$ disappears from the equations and $v$ decouples.
Similarly, since they correspond to $v=0$, that is to
$\phi\rightarrow\infty$, their coordinates themselves do
not depend on $n$ (in other words for $\phi\rightarrow\infty$ all exponents
$n\neq0$, of the same sign, are equivalent).

\subsubsection{Phase-space analysis at infinity}

Due to the fact that the dynamical system
\eqref{eqpowerx}-\eqref{eqpowerw} is non-compact, there could be
features in the asymptotic regime which are non-trivial for the
global dynamics. Thus, in order to complete the analysis of the
phase space we must extend our study using the Poincar\'e
central projection method \cite{PoincareProj}.

We consider the Poincar\'e variables
\begin{equation}\label{Transf0}
x_r=\rho  \cos \theta  \sin \Phi  \sin \psi ,y_r=\rho  \cos \psi, z_r=\rho
\sin \theta  \sin\Phi
    \sin \psi,v_r=\rho  \cos \Phi \sin \psi,
\end{equation}
where $\rho=\frac{r}{\sqrt{1+r^2}},$ $r=\sqrt{x^2+y^2+z^2+v^2},$
$\theta,\Phi\in[0,2\pi],$
and $-\frac{\pi}{2}\leq \psi \leq \frac{\pi}{2}$ (we
restrict the angle $\psi$ to this interval since the physical
region is given by $y>0$)
\cite{PoincareProj,Leon:2008de,Xu:2012jf,Leon2011}. Thus, the points at
infinity
($r\rightarrow+\infty$) are those having $\rho\rightarrow 1$. Furthermore,
the physical phase space is given by
$\left(x_r,y_r,z_r,v_r\right)\in [-1,1]\times [0,1]\times[-1,1]\times
[-1,1]$, such that
\begin{equation}
\frac{x_r^2+y_r^2}{1-v_r^2-x_r^2-y_r^2-z_r^2}-\frac{x_r^2 z_r
\left(\sqrt{6} n v_r x_r-6 v_r^2-6
   x_r^2-6 y_r^2-6
z_r^2+6\right)}{\left(1-v_r^2-x_r^2-y_r^2-z_r^2\right)^{5/2}}\leq 1,
\end{equation}
and $v_r^2+x_r^2+y_r^2+z_r^2\leq 1.$

Inverting relations (\ref{Transf0}) and substituting into
(\ref{Omegaspowerlaw}),(\ref{wdephasepowerlaw}), we obtain the dark energy
density and
equation-of-state parameters as a function of the  Poincar\'e variables,
namely:
\begin{eqnarray}
\Omega_{DE}=\frac{x_r^2+y_r^2}{1-v_r^2-x_r^2-y_r^2-z_r^2}-\frac{x_r^2 z_r
\left(\sqrt{6} n v_r x_r-6 v_r^2-6
   x_r^2-6 y_r^2-6
z_r^2+6\right)}{\left(1-v_r^2-x_r^2-y_r^2-z_r^2\right)^{5/2}},
 \label{Omegas22powerlaw}
 \end{eqnarray}
 \begin{eqnarray}
&&w_{DE}=
\left\{\left[2 z_r \sqrt{\zeta } \left(\sqrt{6} n v_r
x_r-3 \zeta \right)+9 x_r^2 z_r^2+\zeta ^2\right] \left[\sqrt{6} n
   v_r x_r^3 z_r+\zeta ^{3/2} \left(x_r^2+y_r^2\right)-6 x_r^2 z_r \zeta
\right]\right\}^{-1}\nonumber\\
&&\ \ \ \ \ \ \ \ \ \ \ \ \ \   \ \
 \sqrt{\zeta } \left[6 n (n+1) v_r^2 x_r^4
z_r^2-6
\sqrt{6} n v_r x_r^3 z_r^2 \zeta +\sqrt{6} n v_r x_r z_r \zeta ^{3/2}
\left(3 x_r^2-2
   y_r^2\right)\right.\nonumber\\
&&\left.\ \ \ \ \ \ \ \ \ \ \ \ \ \ \ \  \ \ \
\ \ \ \ \ \  +6 z_r \zeta ^{5/2} \left(y_r^2-2 x_r^2\right)+x_r z_r
\zeta ^2 \left(9 x_r z_r-\sqrt{6} y_r^2 \lambda \right)+\zeta ^3
   (x_r^2-y_r^2) \right],
\label{wdephase22powerlaw}
\end{eqnarray}
where $\zeta=1-x_r^2-y_r^2-z_r^2-v_r^2$,
and similarly substituting into
(\ref{wtotpowerlaw}), (\ref{deccpowerlaw})  we obtain the corresponding
expressions for
the total equation-of-state and deceleration parameters:
\begin{eqnarray}
&&w_{tot}
=\left[\zeta^2\left(2 \sqrt{6}\sqrt{\zeta} n v_r x_r z_r  +9 x_r^2
z_r^2-6
z_r
\zeta ^{3/2}+\zeta ^2\right)\right]^{-1}\left[6 n (n+1) v_r^2 x_r^4
z_r^2\right.\nonumber\\
&& \ \ \ \ \ \ \   \  \  -6
\sqrt{6} n v_r x_r^3 z_r^2 \zeta
+\sqrt{6} n v_r x_r z_r \zeta ^{3/2} \left(3 x_r^2-2 y_r^2\right)+6
   z_r \zeta ^{5/2} \left(y_r^2-2 x_r^2\right)\nonumber\\
&&  \ \ \ \ \ \ \ \ \ \ \ \ \ \ \  \ \ \ \ \ \ \ \ \ \ \ \ \ \ \  \  \ \ \
\ \ \ \ \ \  \  \ \left.+x_r z_r \zeta ^2 \left(9
x_r z_r-\sqrt{6} y_r^2 \lambda \right)+\zeta ^3 (x_r^2-y_r^2)
   \right],
\label{wtotphase22powerlaw}
\end{eqnarray}
\begin{eqnarray}
&&q=\left\{2 \zeta ^2 \left[2 z_r \sqrt{\zeta } \left(\sqrt{6} n v_r x_r-3
\zeta \right)+9 x_r^2 z_r^2+\zeta ^2\right]\right\}^{-1}\left\{18 n (n+1)
v_r^2 x_r^4 z_r^2\right. \nonumber\\
 &&  \ \ \ \ \ \ \ \ \  +9 \sqrt{6} n v_r x_r^3 z_r
\zeta  \left(\sqrt{\zeta }-2 z_r\right)+\sqrt{6} x_r z_r \zeta ^{3/2}
\left[2 n v_r
   \left(\zeta -3 y_r^2\right)-3 y_r^2 \sqrt{\zeta } \lambda
\right] \nonumber\\
&& \ \ \ \ \ \ \ \ \  \ \ \ \ \ \ \ \ \ \ \ \left. +3 x_r^2 \zeta ^2
\left(12 z_r^2-12
z_r \sqrt{\zeta }+\zeta \right)+\zeta ^{5/2} \left(\zeta -3
   y_r^2\right) \left(\sqrt{\zeta }-6 z_r\right)\right\}.
\label{wqphase22powerlaw}
\end{eqnarray}
Finally, substitution into (\ref{cs2powerlaw}),(\ref{QSpowerlaw})
provides the corresponding expressions for the instability-related
quantities:
\begin{eqnarray}
&&c_S^2=\left\{x_r \left[2 z_r \sqrt{\zeta } \left(\sqrt{6} n v_r x_r-3
\zeta \right)+9 x_r^2
   z_r^2+\zeta ^2\right]^2\right\}^{-1}
\left\{2 \sqrt{6} n v_r x_r^2
   z_r \zeta ^2 \left(\sqrt{\zeta }-4 z_r\right)\right. \nonumber\\
&&  \ \ \ \ \  \ \ \ \ \ \ \ \ \ 3 x_r^3 z_r^2 \zeta  \left[2
(n-1) n v_r^2-6 z_r \sqrt{\zeta }+5 \zeta \right] +3 \sqrt{6} n v_r x_r^4
z_r^3 \sqrt{\zeta }-27 x_r^5 z_r^4
\nonumber\\
 &&  \ \ \ \ \  \ \ \ \  \    \ \ \ \  \ \ \ \ \ \   \ \ \ \
\  \ \ \ \ \ \ \left. +x_r \zeta ^2 \left[\zeta \left(15 z_r^2-4 z_r
\sqrt{\zeta }+\zeta \right)-9 y_r^2
   z_r^2\right]+\sqrt{6} y_r^2 z_r \zeta ^3 \lambda\right\},
\label{cspoinc1powerlaw}
\end{eqnarray}
\begin{eqnarray}
Q_S=\frac{3 x_r^2 \left[2 z_r \sqrt{\zeta } \left(\sqrt{6} n v_r x_r-3
\zeta \right)+9 x_r^2
   z_r^2+\zeta ^2\right]}{\left(3 x_r^2 z_r+\zeta ^{3/2}\right)^2}.
\label{Qspoinc1powerlaw}
\end{eqnarray}

\begin{table*}[th!]
\begin{center}
\begin{tabular}{|c|c|c|c|c|c|c|c|c|c|c|c|}
\hline
 Cr. P. & $x_r$ & $y_r$ & $z_r$ & $v_r$ &Stability&  $\Omega_{DE}$ &
$w_{DE}$ &
$w_{tot}$ & q & $c_S^2$ &
$Q_S$  \\
\hline \hline
 $L^\pm$&  $0$ & $0$ & $\pm 1$& 0 & unstable & $0$ & arbitrary &
$0$ &
$\frac{1}{2}$ & arbitrary & $0$\\
\hline
 $M^\pm$&  $0$ & $0$ & 0 & $\pm 1$& saddle point & 0&1&0&$\frac{1}{2}$& 1&0
\\
\hline
\end{tabular}
\end{center}
\caption[crit1]{\label{critmod2inf} Scenario 2: Exponential potential and
power-law coupling
  function. The real and physically meaningful
critical points at infinity of the autonomous system
\eqref{eqpowerx}-\eqref{eqpowerw}, and the
corresponding values of the dark-energy
density parameter $\Omega_{DE}$, of the  dark-energy equation-of-state
parameter $w_{DE}$, of the total equation-of-state parameter
  $w_{tot}$ and of the deceleration parameter $q$. The last two columns
contain the instability-related parameters $c_S^2$ and $Q_S$ (from
(\ref{cspoinc1powerlaw}),(\ref{Qspoinc1powerlaw})), which must respectively
be non-negative and
positive  for a scenario free of ghosts and instabilities.
}
\end{table*}
Therefore, performing the analysis described in Appendix \ref{Poincare2}
we
are led to the critical  points at infinity  $L^\pm$ and $M^\pm$, which are
presented in Table \ref{critmod2inf} along with their stability
conditions.

\section{Cosmological Implications}
\label{implications}

Since we have performed the complete phase-space analysis of a quite
general subclass of generalized Galileon scenario, we can now discuss the
corresponding cosmological behavior. In the following subsections we
analyze the two different scenarios, namely that of exponential potential
and exponential coupling function, and that of exponential potential and
power-law coupling function, separately. In each scenario we analyze the
observables at the solutions, and moreover the
instability-related quantities $c_S$ and $Q_S$, examining whether they
satisfy (\ref{cscon}),(\ref{Qscon}) in order for the corresponding
solution to be physical. However, we mention that this ghost investigation
is applied at the critical points, thus even if they are found to be
ghost and instabilities free, it is not a proof that the universe evolution
towards them did not pass through a state with ghosts and Laplacian
instabilities. Therefore in order to obtain a physical evolution of the
universe one has to  ensure that conditions (\ref{cscon}),(\ref{Qscon})
are satisfied everywhere in the examined evolution, and not only at late
times.

\subsection{Scenario 1: Exponential potential and exponential coupling
  function}

The critical points of the scenario at hand are the following.
Points $A^\pm$ exist always, that is for every values of the
scenario parameters $\lambda_V$,$\lambda_g$,$V_0$,$g_0$, they are unstable
or saddle, and thus they cannot be the late-time state of the universe.
They correspond to a non-accelerating, dark-energy dominated universe, with
a stiff dark-energy equation-of-state parameter equal to 1. Finally,
the instability-related quantities $c_S$ and $Q_S$ do satisfy the
corresponding conditions (\ref{cscon}),(\ref{Qscon}), namely $c_S\geq0$ and
$Q_S>0$, and thus these solutions are free of instabilities. Both of them
exist in standard quintessence \cite{Copeland:1997et}.

Point $O_1$ is a saddle one and thus it cannot attract the universe at
late times. It corresponds to a non-accelerating, dark-matter dominated
universe, with  zero total equation-of-state parameter. The
instability-related quantities $c_S$ and $Q_S$ satisfy
(\ref{cscon}),(\ref{Qscon}), and thus this solution is free of
instabilities (the fact that $Q_S$ is exactly 0 and not positive is not
a problem, since this happens only at one point and it is not zero
identically, in which case there might be a non-perturbative ghost
 \cite{DeFelice:2010hg,DeFelice:2011ka}\footnote{We thank A. De Felice
for this comment.}). This point exists in standard quintessence too
\cite{Copeland:1997et}.

Point $C$ exists for $0<\lambda_V^2<6$ and it is a stable one in the region
of the parameter space shown in Table \ref{crit}, and thus it can attract
the universe at late times. It corresponds to a dark-energy dominated
universe, with a dark-energy equation-of-state parameter lying in the
quintessence regime, which can be accelerating or not according to the
$\lambda_V$-value. Additionally, this solution is free of instabilities.
This point exists in standard quintessence \cite{Copeland:1997et}. It is
quite important, since it is both stable and possesses $w_{DE}$ and $q$
compatible with observations.

Furthermore, we mention that in the specific case where $\lambda_V=0$, that
is in the case of constant or zero usual potential, there exist the
critical point $C_0$, and it is always stable. It corresponds to the de
Sitter solution, where the universe is accelerating and  dark-energy
dominated, with the dark-energy behaving like a cosmological constant
($w_{DE}=-1$), and it is free of instabilities. This point exists in
standard quintessence \cite{Copeland:1997et}. It is  quite important, since
it is both stable and possesses $w_{DE}$ and $q$ compatible with
observations. Although at first sight it seems to be the
$\lambda_V\rightarrow0$ limit of point $C$, this is not the case since the
eigenvalues of $C_0$ do not arise from the $\lambda_V\rightarrow0$ limit of
$C$-eigenvalues ($\lambda_V=0$ is a bifurcation value) and thus it has to
be considered separately. Moreover, since in many Galileon works the
authors do
not consider a potential,  point $C_0$ just gives the late-time state of
the universe in these cases \cite{DeFelice:2011bh,Appleby:2011aa}.

Point $D$ exists for $\lambda_V^2\geq3$ and in this case it is always
stable, that is it can be the late-time state of the universe, and it is
free of instabilities. It has the advantage that the dark-energy density
parameter is in the interval $0<\Omega_{DE}<1$, that is it can alleviate
the coincidence problem since dark energy and dark matter density
parameters can be of the same order. However, it has the disadvantage that
it is not accelerating and $w_{DE}=0$, which are not favored by
observations. This point exists in standard quintessence
\cite{Copeland:1997et} too.
 \begin{figure}[ht]
\begin{center}
\includegraphics[width=8.1cm,height=8.1cm,clip=true]{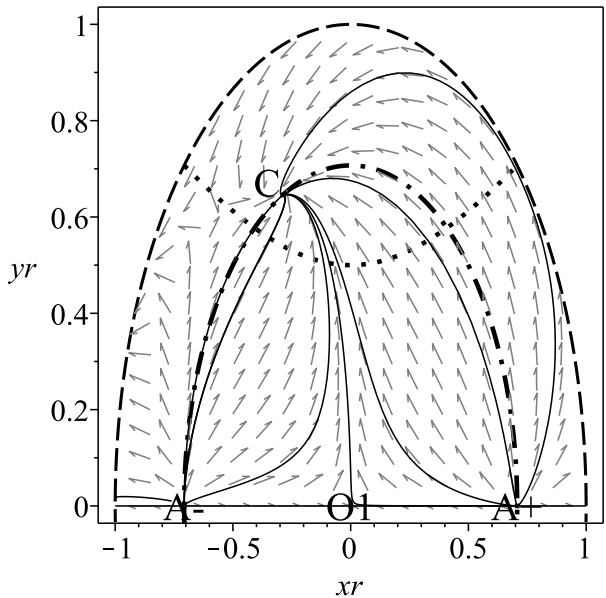}
\caption{{\it{
Trajectories in the $y_r$-$x_r$ plane of the Poincar\'e phase space  for
the Scenario 1, that is for exponential potential and exponential
coupling function. We use $\lambda_V=1$ and $\lambda_g$
arbitrary (for the numerics we choose
$\lambda_g=1$ but different
$\lambda_g$'s correspond to the same projection on $y_r$-$x_r$ plane). The
region inside the inner semi-circle (seen as semi-ellipse in the figure
scale), marked by the
thick dashed-dotted line, is the
physical part of the phase space. The region above the dotted line marks
the region corresponding to accelerating universe ($q<0$). In
this projection the dark-energy dominated, accelerating, quintessence-like
solution
$C$ is a
stable solution, $O_1$ is saddle point, and $A^\pm$ are unstable.}}}
\label{fig1}
\end{center}
\end{figure}

Apart from the above points that exist also in standard quintessence, the
scenario at hand possesses two additional critical points, namely
$B^\pm$. They correspond to dark-energy domination, with a dark-energy
equation-of-state parameter lying in the quintessence regime, where the
universe is non-accelerating ($q>0$), and they are free of instabilities.
However, these points are not stable and thus they cannot attract the
universe at late times.

Finally, the present scenario possesses two critical point at infinity,
namely $K^\pm$. They correspond to a dark-matter dominated,
non-accelerating universe, with arbitrary $w_{DE}$ but with a zero total
equation-of-state parameter $w_{tot}$, which are also free of
instabilities. They are always unstable and therefore they cannot be the
late-time state of the universe.

 From the above analysis we observe that at the stable critical points,
 $C$ and $D$, we have $\dot{\phi}\rightarrow0$,
$\phi\rightarrow-\text{sign}(\lambda_V)\infty$ and thus for
$\lambda_V\lambda_g>0$ we obtain $g(\phi)\rightarrow0$ while for
$\lambda_V\lambda_g<0$ we obtain $g(\phi)\rightarrow\infty$ (for
$\lambda_V=0,$ $g(\phi)$ can be zero, finite, or infinity). Similarly,
for $C_0$ we see that  for
$\lambda_g>0$ we obtain $g(\phi)\rightarrow0$ while for
$\lambda_g<0$ we obtain $g(\phi)\rightarrow\infty$. In all cases, if
$\lambda_g=0$ then obviously $g(\phi)=const$. From these we deduce that
Galileons (simple or generalized) may survive at late-time cosmology or
may be completely disappeared by the dynamics, depending on the model
parameters. However, firstly we observe that Galileon cosmology possesses
the same stable late-time solutions with standard quintessence
\cite{Copeland:1997et}. Secondly, and more interestingly,
as we observe from Table \ref{crit1}, the corresponding observables of
these solutions do not depend on the Galileon terms, but only on the usual
terms and especially on the standard scalar potential (note that even the
instability-related quantities do not depend on the Galileon terms either).
This is a main result of the present work.
 \begin{figure}[ht]
\begin{center}
\includegraphics[width=8.1cm,height=8.1cm,clip=true]{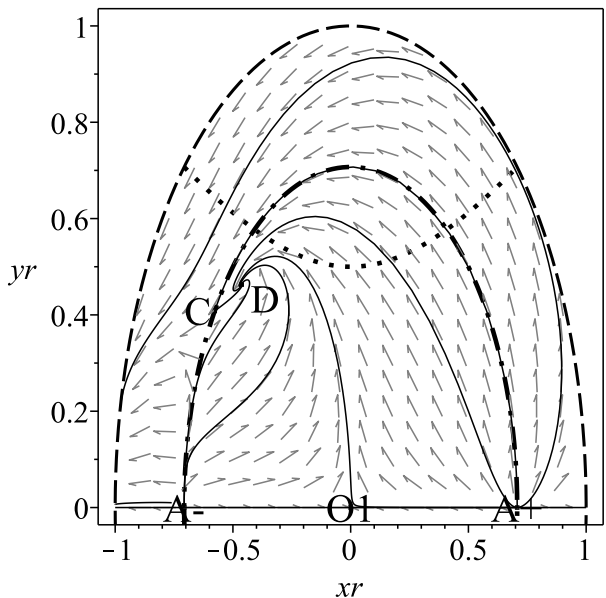}
\caption{{\it{
Trajectories in the $y_r$-$x_r$ plane of the Poincar\'e phase space  for
the Scenario 1, that is for exponential potential and exponential
coupling function. We use  $\lambda_V=2$ and $\lambda_g$
arbitrary (for the numerics we choose
$\lambda_g=1$ but different
$\lambda_g$'s correspond to the same projection on $y_r$-$x_r$ plane). The
region inside the inner semi-circle (seen as semi-ellipse in the figure
scale), marked by the
thick dashed-dotted line, is the
physical part of the phase space. The region above the dotted line marks
the region corresponding to accelerating universe ($q<0$). In
this projection the non-accelerating, dust-like ($w_{DE}=0$) solution $D$
is a stable spiral, $C$ and $O_1$ are saddle points, and
$A^\pm$ are
unstable.}}}
\label{fig2}
\end{center}
\end{figure}

Therefore, although
the Galileon terms of generalized Galileon cosmology can play an important
role at early (inflationary) times
\cite{Creminelli:2010ba,Kobayashi:2010cm,Burrage:2010cu,
Liu:2011ns,Kobayashi:2011nu,Ohashi:2012wf,Choudhury:2012yh,
Hinterbichler:2012fr} or at recent
times
\cite{Silva:2009km,Gannouji:2010au,DeFelice:2010pv,Mota:2010bs,
DeFelice:2011bh,
Shirai:2012iw,Tretyakov:2012zz,Germani:2012qm,Sampurnanand:2012wy,
Zabat:2011zz}, in the
future they will not have any effect on the universe evolution, when it
will asymptotically reach its stable state. This is consistent with
observational constraints, which disfavor the Galileon presence in the
current universe
\cite{Kobayashi:2009wr,Ali:2010gr,Nesseris:2010pc,Hirano:2011wj,
Appleby:2012ba,Ali:2012cv} (see also \cite{Hossain:2012qm}). Finally, one
could still ask whether going beyond the background analysis, and examining
perturbation-related observables at the late-time stable solutions, would
reveal a dependence of these observables on the Galileons. Although this
complicated project lies beyond the scope of the present work, the
examination of the instability-related quantities $c_S$ and $Q_S$, which
arise from a perturbative analysis, and their independence on the Galileon
terms, is an indication that even the perturbation observables will not
depend on the Galileon terms.
  \begin{figure}[ht]
\begin{center}
\includegraphics[width=8.1cm,height=8.1cm,clip=true]{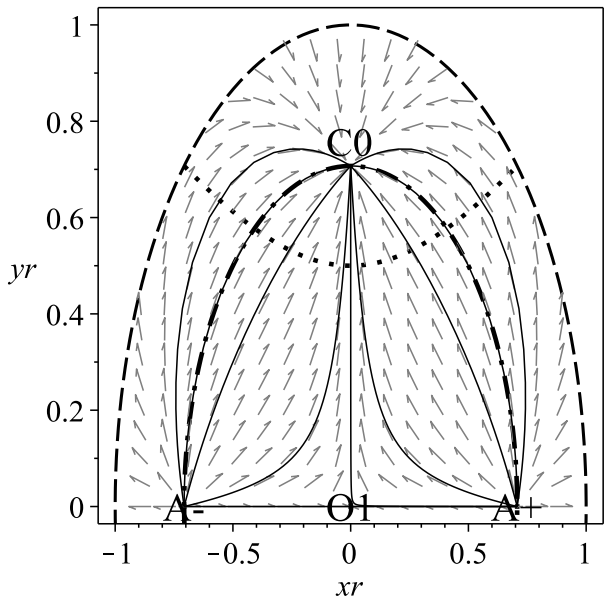}
\caption{{\it{
Trajectories in the $y_r$-$x_r$ plane of the Poincar\'e phase space  for
the Scenario 1, that is for exponential potential and exponential
coupling function, for the specific case $\lambda_V=0$ and $\lambda_g$
arbitrary (for the numerics we choose
$\lambda_g=1$ but different
$\lambda_g$'s correspond to the same projection on $y_r$-$x_r$ plane). The
region inside the inner semi-circle (seen as semi-ellipse in the figure
scale), marked by the
thick dashed-dotted line, is the
physical part of the phase space. The region above the dotted line marks
the region corresponding to accelerating universe ($q<0$). In this
projection the de Sitter solution
$C_0$ is a stable node,  $O_1$ is saddle point, and $A^\pm$ are
unstable.}}}
\label{fig3}
\end{center}
\end{figure}

In order to present the above behavior in a more transparent way, we
first evolve the autonomous system \eqref{eqx}-\eqref{eqz}
numerically for the choice $\lambda_V=1$ and $\lambda_g$
arbitrary (with $\lambda_g>-\lambda_V$), and
in Fig. \ref{fig1} we depict the corresponding phase-space behavior,
projected in the Poincar\'e $x_r$-$y_r$ plane (for the numerics we choose
$\lambda_g=1$ but different
$\lambda_g$'s correspond to the same projection on $y_r$-$x_r$ plane). As
we can see, in this case the dark-energy dominated, accelerating,
quintessence-like critical point  $C$ is the
late-time solution of the universe.

Similarly, in  Fig. \ref{fig2} we depict the corresponding phase-space
behavior, for  $\lambda_V=2$ and $\lambda_g$ arbitrary (we use
$\lambda_g=1$, which satisfies $\lambda_g>-\lambda_V$). In this case the
non-accelerating critical point  $D$ is the late-time solution of the
universe, while $C$ is saddle point. Finally, in  Fig. \ref{fig3} we depict
the phase-space behavior for the
specific case $\lambda_V=0$ and $\lambda_g$
arbitrary. In this case the universe is attracted by the de Sitter
solution $C_0$ as expected.

\subsection{Scenario 2: Exponential potential and power-law coupling
  function}

This scenario possesses the following critical points. $E^\pm$ exist
for every value of the parameters $\lambda_V$,$n$,$V_0$,$g_0$, they are
saddle points, and thus they cannot be the late-time state of the universe.
They correspond to a non-accelerating, dark-energy dominated universe, with
a stiff dark-energy equation-of-state parameter equal to 1, and they are
free of instabilities. Both of them exist in standard quintessence
\cite{Copeland:1997et}.

Point $O_2$ is a saddle one and thus it cannot attract the universe at
late times. It corresponds to a non-accelerating, dark-matter dominated
universe, with zero total equation-of-state parameter, and it
is free of instabilities. This point exists in standard quintessence too
\cite{Copeland:1997et}.

Point $F$ exists for $0<\lambda_V^2<6$ and it is a stable one for
$\lambda_V^2<3$  and thus it can be the late-time state of the universe.
It corresponds to a dark-energy dominated universe, with a dark-energy
equation-of-state parameter in the quintessence regime, which can be
accelerating or not according to the $\lambda_V$-value. Additionally, this
solution is free of instabilities. This point is quite important, since it
is both stable and possesses $w_{DE}$ and $q$ compatible with
observations. It exists in standard quintessence \cite{Copeland:1997et}
too.

There is another interesting critical point, namely $F_0$, which is
obtained only in the case where $\lambda_V=0$, that is in the case of
constant or zero usual potential, and it is always stable. It corresponds
to the de Sitter solution, where the universe is accelerating and
dark-energy dominated, with the dark-energy behaving like a cosmological
constant ($w_{DE}=-1$), and it is free of instabilities. At first sight it
seems to be the $\lambda_V\rightarrow0$ limit of point $F$, however this is
not the case since the eigenvalues of $F_0$ do not arise from the
$\lambda_V\rightarrow0$ limit of $F$-eigenvalues ($\lambda_V=0$ is a
bifurcation value) and thus it has to be considered separately.
Furthermore, since in many Galileon works the standard potential is not
considered, point $F_0$ is straightforwardly the corresponding late-time
state of the universe in these cases.

Point $G$ exists for $\lambda_V^2\geq3$ and in this case it is always
stable, that is it can attract the universe at late times, and it is free
of instabilities. It has the advantage that the dark-energy density
parameter lies in the interval $0<\Omega_{DE}<1$, that is it can alleviate
the coincidence problem, but it has the disadvantage that it is not
accelerating and possesses $w_{DE}=0$, which are not favored by
observations. This
point exists in standard quintessence \cite{Copeland:1997et} too.

Finally, the scenario at hand possesses the critical point at infinity
$L^\pm$ and $M^\pm$. They correspond to a dark-matter dominated,
non-accelerating universe, with a zero total
equation-of-state parameter $w_{tot}$, which are also free of
instabilities. $M^\pm $ are saddle points whereas $L^\pm$ are unstable, and
thus they cannot be the late-time state of the universe.

>From the above analysis we can see that at the stable critical points,
 $F$ and $G$, we have $\dot{\phi}\rightarrow0$,
$\phi\rightarrow-\text{sign}(\lambda_V)\infty$ and thus for
$n<0$ we obtain $g(\phi)\rightarrow0$ while for
$n>0$ we obtain $g(\phi)\rightarrow\infty$. Similarly,
for $F_0$ we see that  for
$n<0$ we obtain $g(\phi)\rightarrow0$ while for
$n>0$ we obtain $g(\phi)\rightarrow\infty$. In all cases, if
$n=0$ then obviously $g(\phi)=const$. From these it is implied that
Galileons (simple or generalized) may survive at late-time cosmology or
may be completely disappeared by the dynamics, depending on the model
parameters. Similarly to the previous subsection, firstly we
observe that this scenario possesses the same stable late-time
solutions with standard quintessence \cite{Copeland:1997et}.
Furthermore, as we observe from Table \ref{crit22}, the
corresponding observables of these
solutions do not depend on the Galileon terms, but only on the usual terms
and especially on the standard scalar potential. Even the
instability-related quantities do not depend on the Galileon terms either.
This is a main result of the present work.

 \begin{figure}[ht]
\begin{center}
\includegraphics[width=10cm,angle=0,clip=true]{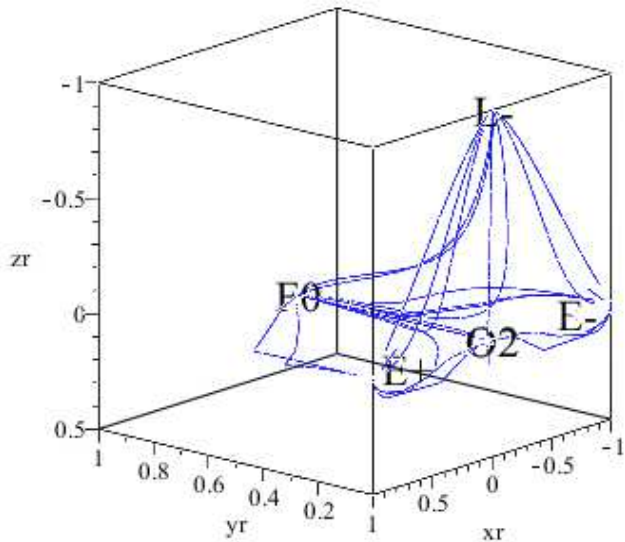}
\caption{{\it{ Projection of   orbits on the $x_r$-$y_r$-$z_r$ space of
the Poincar\'e phase space  for
the Scenario 2, that is for exponential potential and power-law
coupling function, for the specific case $\lambda_V=0$ and $n$
arbitrary (for the numerics we choose $n=1$ but different
$n$'s correspond to the same projection). In this projection the de Sitter
solution $F_0$ is the attractor, whereas $L^-$ and  $E^\pm$ are unstable
and $O_2$ is a
saddle point.}}}
\label{fig8}
\end{center}
\end{figure}
Thus, and similarly to the scenario of the previous
subsection, although
the Galileons can play an important
role at  inflationary
\cite{Creminelli:2010ba,Kobayashi:2010cm,Burrage:2010cu,
Liu:2011ns,Kobayashi:2011nu,Ohashi:2012wf,Choudhury:2012yh,
Hinterbichler:2012fr} or at recent
times
\cite{Silva:2009km,Gannouji:2010au,DeFelice:2010pv,Mota:2010bs,
DeFelice:2011bh,
Shirai:2012iw,Tretyakov:2012zz,Germani:2012qm,Sampurnanand:2012wy,
Zabat:2011zz}, in the
future, when the universe
will asymptotically reach its stable state, they will not have any effect
on its evolution (this is consistent with
observational constraints which disfavor the Galileon presence in the
current universe
\cite{Kobayashi:2009wr,Ali:2010gr,Nesseris:2010pc,Hirano:2011wj,
Appleby:2012ba,Ali:2012cv}. One could still ask whether the observables
related to perturbations would depend on the Galileon terms, however the
investigation of the instability-related quantities $c_S$ and $Q_S$, which
come from a perturbative analysis, and their independence on the Galileon
terms, is an indication that even the perturbation observables will not
depend on the Galileon terms.

In order to present the above behavior in a more transparent way, we
first evolve the autonomous system \eqref{eqpowerx}-\eqref{eqpowerw}
numerically for the choice $\lambda_V=0 $ and $n$
arbitrary, and
in Fig. \ref{fig8} we depict the corresponding phase-space behavior,
projected on the Poincar\'e $x_r$-$y_r$-$z_r$ space (for the numerics we
choose
$n=1$ but different
$n$'s correspond to the same projection). As
we can wee, in this case the de Sitter critical point $F_0$ is the
late-time solution of the universe while $L^-$ and  $E^\pm$ are unstable.
Finally, note that the sign of $z_r$ is invariant, thus the
plane $z_r=0$ cannot be crossed by the orbits.

We close this section by comparing the present scenario of exponential
potential and power-law coupling function, with the one of the previous
subsection, that is with exponential potential and exponential coupling
function. A first observation is that for $\lambda_g=0$ and $n=0$ the
critical points of both scenarios coincide, which was expected since in
this case the two models coincide and they both transit to the simple
Galileon scenario, where the couplings are constants and not functions of
the scalar field. Furthermore, in both scenarios we found that there are
not additional stable late-time solutions apart from those of standard
quintessence. What is more interesting is that all the observables do not
depend on the Galileon terms, that is the Galileons do not play any role at
late times. However, in general the two scenarios are different, with the
main difference being that they have different non-stable critical points
and thus phase-space dynamics.

\section{Conclusions}
\label{Conclusions}

In the present work we have investigated the dynamical behavior of generalized
Galileon cosmology, which is a recent construction based on higher
derivatives in the action, along with the imposition of the Galilean
symmetry, with the requirement of maintaining the equations of motion
second ordered. However, one additionally extends the constant coefficients
of the various action-terms of the simple Galileon formulation, into
arbitrary functions of the scalar field.

Performing a detailed phase-space analysis of two distinct and quite
general scenarios, namely of exponential usual potential and exponential
Galileon coupling-function, and exponential usual potential and  power-law
Galileon coupling-function, we extracted the stable solutions, that is the
solutions that will be the late-time states of the universe. In each of
these stable late-time solutions we have calculated various observables, such as
the dark-energy equation-of-state parameter, the deceleration parameter,
and the dark-energy and matter density parameters. Furthermore, in order to
examine whether these solutions are free of ghosts and instabilities, we
calculate the corresponding instability-related quantities of the
literature.

In the case where the Galileon terms are absent we recovered the results of
standard quintessence. In the case where the Galileons are present we
found that at late times they may survive at or they may   completely
disappear by the dynamics, depending on the model parameters. However,
independently of their disappearance or survival, we found that the
scenario at hand possesses exactly the same stable late-time solutions with
standard quintessence, which are moreover free of ghosts and instabilities.
More interestingly, the corresponding observables at these stable late-time
solutions do not depend on the Galileon terms, but only on the usual
action-terms and especially on the standard scalar potential. This is a
main result of the present work, and it shows that  although
the Galileons can play an important role at  inflationary
\cite{Creminelli:2010ba,Kobayashi:2010cm,Burrage:2010cu,
Liu:2011ns,Kobayashi:2011nu,Ohashi:2012wf,Choudhury:2012yh,
Hinterbichler:2012fr}
 or at recent
times \cite{Silva:2009km,Gannouji:2010au,DeFelice:2010pv,Mota:2010bs,
DeFelice:2011bh,
Shirai:2012iw,Tretyakov:2012zz,Germani:2012qm,Sampurnanand:2012wy,
Zabat:2011zz}, in the
future,  when the universe will asymptotically reach its stable state, they
will not have any effect on its evolution. This is in agreement with
observational constraints which disfavor the Galileon presence in the
current universe
\cite{Kobayashi:2009wr,Ali:2010gr,Nesseris:2010pc,Hirano:2011wj,
Appleby:2012ba,Ali:2012cv}.

One could ask whether the above behavior is a result of the specific
ansantzes, or it has a general character. Although one cannot exclude the
case where suitably chosen or tuned ansatzes can lead to significant
Galileon effects on the observables at late times, in general one expects
the above behavior to be valid at late times in the large majority of
cases, due to the downgrading of the role of the derivative (Galileon)
terms in an eternally expanding universe. However,   we mention
that in the present work we followed the majority of the
Galileon cosmological works and we viewed the Galileon theory as a
scalar-field, dark-energy, construction, and not as a modified gravity,
and thus we did not include a coupling between the scalar field and the
matter sector which existed in the initial Galileon formulation. Such a
coupling could indeed lead to significantly different cosmological
behavior, and its investigation is left for a future project.
   
We close this work by mentioning that the obtained result that the
Galileons will not play any role and that they will not have any
observable effect at late times, was extracted for the background
evolution and the corresponding observables. There could still be the case
that Galileons could leave their signatures in observables
related to perturbations. Although such an investigation is quite
complicated and lies beyond the scope of the present work, the examination
of instability-related quantities arising from perturbation analysis that
we did perform in this work showed that they do not depend on the Galileon
terms either, which is a quite strong indication that even the perturbation
observables will not depend on the Galileon terms.

 \begin{acknowledgments}
The authors wish to thank S. Lepe for fruitful discussions and S. A.
Appleby, P. Creminelli, A. De Felice, J. Khoury, E. V. Linder  and A.
Nicolis, for useful comments. The research project is implemented within
the framework of the Action ``Supporting Postdoctoral Researchers'' of the
Operational Program ``Education and Lifelong Learning'' (Actions
Beneficiary: General Secretariat for Research and Technology), and is
co-financed by the European Social Fund (ESF) and the Greek State.
Additionally, it is co-financed by MECESUP FSM0806 from Ministerio de
Educaci\'on,
Chile and by PUCV through Proyecto DI Postdoctorado 2013. GL wishes to
thank to his colleagues
at Instituto de F\'{\i}sica, Pontificia Universidad
de Cat\'{o}lica de Valpara\'{\i}so for their warm hospitality during the
completion of this work.
 \end{acknowledgments}

\begin{appendix}

\section{Stability of Scenario 1}

\subsection{Stability of the finite critical points}
\label{appmod1a}

For the critical points $(x_c,y_c,z_c)$ of the autonomous system
system \eqref{eqx}-\eqref{eqz}, the coefficients of the perturbation
equations form a $3\times3$ matrix ${\bf {Q}}$, which can be easily
obtained, however, since they are rather lengthy expressions    we
do not present them explicitly. Despite this complicated form, using
the specific critical points presented in Table \ref{crit} the matrix
${\bf {Q}}$ acquires a simple form that allows for an easy calculation of
its eigenvalues. The corresponding eigenvalues $\nu_i$ ($i=1,2,3$)
for each critical point are presented in Table
\ref{eigen1}.
\begin{table*}[!]
\begin{center}
\begin{tabular}{|c|c|c|c|}
\hline
&&&  \\
 Cr. P.&  $\nu_1$ &  $\nu_2$ & $\nu_3$  \\
\hline \hline
&&&   \\
$A^+$& $3$ & $\sqrt{6} \lambda_g-6$ & $3+\sqrt{\frac{3}{2}} \lambda_V$
\\
\hline
&&&   \\
$A^-$& $3$ & $-\sqrt{6} \lambda_g-6$ & $3-\sqrt{\frac{3}{2}} \lambda_V$
\\
\hline
&&&   \\
 $B^+$& $\frac{3}{4} \left[-\frac{\sqrt{2} \lambda_g}{\sqrt{\alpha
^-(\lambda_g)-3}}+\alpha ^+(\lambda_g)-4\right]$ & $\frac{3}{4}
\left[\frac{\sqrt{2} \lambda_g}{\sqrt{\alpha ^-(\lambda_g)-3}}+\alpha
   ^+(\lambda_g)-4\right]$ & $\frac{(\lambda_g+\lambda_V) \alpha
^+(\lambda_g)}{2 \lambda
   g}$ \\
\hline
&&&   \\
 $B^-$& $\frac{3}{4} \left[-\frac{\sqrt{2} \lambda_g}{\sqrt{\alpha
^+(\lambda_g)-3}}+\alpha ^-(\lambda_g)-4\right]$ &
$\frac{3}{4} \left[\frac{\sqrt{2} \lambda_g}{\sqrt{\alpha
^+(\lambda_g)-3}}+\alpha
   ^-(\lambda_g)-4\right]$ & $\frac{(\lambda_g+\lambda_V) \alpha
^-(\lambda_g)}{2 \lambda_g}$ \\
\hline
&&&   \\
 $C$& $\lambda_V^2-3$ & $-\lambda_V (\lambda_g+\lambda_V)$ & $\frac{1}{2}
\left(\lambda_V^2-6\right)$ \\
\hline
&&&   \\
 $C_0$& $-3$ & $-3$ & $-3$ \\
\hline
&&&    \\
$D$& $-\frac{3 (\lambda_g+\lambda_V)}{\lambda_V}$ & $\frac{3}{4}
\left(-\frac{\sqrt{24 \lambda_V^2-7 \lambda_V^4}}{\lambda_V^2}-1\right)$ &
$\frac{3}{4} \left(\frac{\sqrt{24 \lambda_V^2-7
   \lambda_V^4}}{\lambda_V^2}-1\right)$ \\
\hline
&&&    \\
$O_1$& $-\frac{9}{2}$ & $-\frac{3}{2}$ &
$\frac{3}{2}$ \\
\hline
\end{tabular}
\end{center}
\caption[crit]{\label{eigen1}
The eigenvalues of the matrix
${\bf {Q}}$ of the perturbation equations of the autonomous system
\eqref{eqx}-\eqref{eqz}. We have defined the functions
$\alpha^\pm(\lambda_g)=\lambda_g^2\pm\lambda_g\sqrt{\lambda_g^2-6}.$ }
\end{table*}

Thus, by determining the sign of the real parts of these
eigenvalues, we can classify the corresponding critical point. In
particular, if all the eigenvalues of a critical point have negative real
parts then the corresponding point is stable, if they all have positive
real parts then it is unstable, and if they change sign then it is a
saddle point.

We mention here that in order to examine correctly the stability of point
$C_0$ (since the second eigenvalue vanishes at first place) we introduce
the local coordinates $$\left\{x,y-1,z \right\}=\epsilon \left\{\tilde{x},
\tilde{y},
\tilde{z} \right\}+{\cal O}(\epsilon^2)$$ where $\epsilon$ is a constant
satisfying $\epsilon\ll 1$. Then we obtain the local evolution equations
\begin{align}
&\tilde{x}'=-3 \tilde{x}+\text{h.o.t},\ \
\ \ \ \tilde{y}'=-3\tilde{y}+\text{h.o.t},\ \ \ \ \
\tilde{z}'=-3\tilde{z}+\text{h.o.t},
\end{align}
with $\text{h.o.t}$ denoting ``higher order terms''. From the above linear
approximation we deduce that the eigenvalues of the linearization for point
$C_0$ are all negative and are displayed in Table \ref{eigen1}, thus $C_0$
is a stable node.

Similarly, in order to examine correctly the stability of point $O_1$ we
introduce the local
coordinates $$\left\{x,y,z \right\}=\epsilon \left\{\tilde{x}, \tilde{y},
\tilde{z} \right\}+{\cal O}(\epsilon^2)$$ with $\epsilon\ll 1.$ The local
evolution equations valid around the origin then write as
\begin{align}
&\tilde{x}'=-\frac{3}{2} \tilde{x}+\text{h.o.t},
\ \ \ \ \ \tilde{y}'=\frac{3}{2}\tilde{y}+\text{h.o.t},\ \ \ \
\ \tilde{z}'=-\frac{9}{2}\tilde{z}+\text{h.o.t}.
\end{align}
The eigenvalues of the linearization for point $O_1$ are displayed in Table
\ref{eigen1}, and thus $O_1$ is a saddle point.

\subsection{Stability of the critical points at infinity}\label{Poincare1}

We consider the Poincar\'e variables
\begin{equation}
\label{Transfapp}
x_r=\rho \cos\theta \sin \psi ,\ z_r=\rho
\sin \theta \sin \psi ,\, y_r= \rho \cos \psi,\,
\end{equation} with $\rho=\frac{r}{\sqrt{1+r^2}},$ $r=\sqrt{x^2+y^2+z^2},$
$\theta\in[0,2\pi],$
and $-\frac{\pi}{2}\leq \psi \leq \frac{\pi}{2}$ (we
restrict the  $\psi$ angle to this range since the physical
region is given by $y>0$)
\cite{PoincareProj,Leon:2008de,Xu:2012jf,Leon2011}.   Therefore, the
points at infinity
($r\rightarrow+\infty$) are those with $\rho\rightarrow 1.$
Moreover, the physical phase space is now given by
$\left(x_r,y_r,z_r\right)\in [-1,1]\times [0,1]\times[-1,1]$ such that
\begin{equation}
\frac{x_r^2+y_r^2}{1-x_r^2-y_r^2-z_r^2}-\frac{x_r^2 z_r
\left(\sqrt{6}\lambda_g x_r-6
   x_r^2-6 y_r^2-6
z_r^2+6\right)}{\left(1-x_r^2-y_r^2-z_r^2\right)^{5/2}}\leq 1,
\end{equation}
and $x_r^2+y_r^2+z_r^2\leq 1.$

Performing the transformation \eqref{Transfapp} on the system
(\ref{eqx})-(\ref{eqz}), and taking
the limit $\rho\rightarrow
1$, the leading terms in the resulting system
are
\begin{eqnarray}
&&\rho'\rightarrow -\lambda_g^2 \cos ^2\theta \sin^2\psi \left[\cos
(2 \theta ) \sin^2\psi +\cos^2\psi \right],\label{inftya}\\
&&(1-\rho ^2) \theta'\rightarrow -2 \lambda_g^2 \sin\theta\cos^3\theta
\sin^2\psi,\label{inftyb}\\
&& (1-\rho ^2) \psi'\rightarrow -2 \lambda_g^2 \sin^2\theta\cos
^2\theta\sin^3\psi\cos\psi.\label{inftyc}
\end{eqnarray}
Since the equation for $\rho$ decouples, we only need to investigate the
subsystem of the angular variables. Thus, equating \eqref{inftyb}
and \eqref{inftyc} to zero, we obtain three classes of solutions:
\\
{\bf Class A}: \begin{eqnarray}
 &&x_r=\sin\psi,\ y_r=\cos\psi, \ z_r=0,\label{PointsatinfinityA}
\end{eqnarray}
for some $\psi$. These points   lead to a divergence in
$\Omega_{DE}$ in (\ref{Omegas22}) and thus we deduce that they are
unphysical.
\\
{\bf Class B}: \begin{eqnarray}
 &&x_r=0,\ y_r=\cos\psi, \ z_r=\pm \sin\psi,\label{PointsatinfinityB}
\end{eqnarray} for some $\psi.$ These points correspond to
$\theta=\pm\frac{\pi}{2}.$   By substituting these values for $x_r, y_r, z_r$ in
the evolution equations
for  $x_r', y_r', z_r'$ we obtain that $y_r$ must be zero at a fixed point
at infinity. Thus, we have the points at infinity given by
$K^\pm:\left[x_r=0,
y_r=0, z_r=\pm 1\right].$
\\
{\bf Class C}: \begin{eqnarray}
 &&x_r=0,\
 y_r=1, \ z_r=0.\label{PointsatinfinityC}
\end{eqnarray} This point leads to a divergence in $\Omega_{DE}$
in (\ref{Omegas22}) and thus it is unphysical.

 In summary, the scenario of exponential potential and exponential
coupling function has only two critical points at infinity, namely
$K^\pm:\left[x_r=0,
y_r=0, z_r=\pm 1\right]$. In order to calculate the eigenvalues of the
linearization of the dynamical system for $x_r,y_r,z_r$ around $K^\pm$, we
proceed as follows: having calculated the linear perturbation matrix
$\mathbf{Q}$ we evaluate it at $y_r=0$, then we take the limit
$x_r\rightarrow 0$ and finally we incorporate the lateral limit
$z_r\rightarrow 1^-$ (that is from below) for $K^+$ or $z_r\rightarrow
-1^+$ (that is from above) for $K^-$. Calculating the eigenvalues of the
resulting   matrix we obtain that the eigenvalues of the
linearization for both $K^\pm$ are given by
$\left(3, \frac{15}{4}, \frac{9}{2}\right)$. Therefore, we conclude that
$K^\pm$ are
unstable.

\section{Stability of Scenario 2}

\subsection{Stability of the finite critical points}
\label{appmod2a}

For the critical points $(x_c,y_c,z_c,v_c)$ of the autonomous system
\eqref{eqpowerx}-\eqref{eqpowerw}, the coefficients of the perturbation
equations form a $4\times4$ matrix ${\bf {Q}}$. The explicit expressions
for
these coefficients are quite lengthy and for simplicity we do not present
them
explicitly. However, using the explicit critical points presented in Table
\ref{crit2} we can straightforwardly see that the matrix ${\bf {Q}}$
acquires a simple form that allows for an easy calculation of its
eigenvalues. The corresponding eigenvalues $\nu_i$ ($i=1,2,3,4$) for each
critical point are presented in Table \ref{eigen2}.
\begin{table*}[ht]
\begin{center}
\begin{tabular}{|c|c|c|c|c|}
\hline
&&&&  \\
 Cr. P.&  $\nu_1$ &  $\nu_2$ & $\nu_3$ & $\nu_4$  \\
\hline \hline
&&& &  \\
$E^{\pm}$& $-6$ & $3\pm\frac{\lambda_V\sqrt{6}}{2}$ & $3$& $0$
\\
\hline
&&&   &\\
 $F$& $\lambda_V^2-3$ & $\frac{\lambda_V^2}{2}-3$ & $-\lambda_V^2$ & $0$ \\
\hline
&&&   &\\
 $F_0$& $-3$ & $-3$ & $-3$ & $0$ \\
\hline
&&&   &\\
$G$& $-3$ & $\frac{3}{4} \left(-\frac{\sqrt{24 \lambda_V^2-7
\lambda_V^4}}{\lambda_V^2}-1\right)$ & $\frac{3}{4}
\left(\frac{\sqrt{24 \lambda_V^2-7
   \lambda_V^4}}{\lambda_V^2}-1\right)$ & $0$\\
\hline
$O_2$& $-\frac{9}{2}$ & $-\frac{3}{2}$ &
$\frac{3}{2}$ & 0 \\
\hline
\end{tabular}
\end{center}
\caption[crit]{\label{eigen2} 
The eigenvalues of the matrix
${\bf {Q}}$ of the perturbation equations of the autonomous system
\eqref{eqpowerx}-\eqref{eqpowerw}. }
\end{table*}

Note that the stability of the above points does not
depend on the exponent $n$, since these points have $z_c=0$ in which case
$n$ disappears from the equations and $v$ decouples.
Similarly, since they correspond to $v=0$, that is to
$\phi\rightarrow\infty$, their coordinates themselves do
not depend on $n$ (in other words for $\phi\rightarrow\infty$ all exponents
$n\neq0$ of the same sign are equivalent).

Strictly speaking, since the eigenvalue of the fourth auxiliary variable
$v$ is always zero, the corresponding points are non-hyperbolic, and thus
one should apply the center manifold analysis \cite{normally} in order to
deduce whether it is unstable, saddle or stable.
However, since all the above points have $z_c=0$, $v$ is completely
decoupled from the   autonomous system
\eqref{eqpowerx}-\eqref{eqpowerw}. Therefore, one can examine only the
three eigenvalues $\nu_1$,$\nu_2$,$\nu_3$  displayed in Table \ref{eigen2},
and the results of the stability analysis are presented in Table
\ref{crit2}. Finally, in order to examine correctly the stability of points
$F_0$ and $O_2$ we use the same linear approximation with that in
the end of appendix \ref{appmod1a} (for points $C_0$ and $O_1$), and we
deduce that $F_0$ is a stable node while $O_1$ is saddle, with eigenvalues
depicted in Table \ref{eigen2}.

We end this appendix with the following comment. The
above stability conditions for $F$, $F_0$ and $G$ we extracted using the
fact that the $v$-direction is decoupled. However, as we mentioned the
complete dynamical picture should be obtained after a full center manifold
analysis \cite{normally}, which is the correct way to handle zero
eigenvalues. Therefore, in the following we present this analysis for
completeness. Since compact variables are preferable
to see the subtleties on the dynamics, we perform the calculations using a
Poincar\'e projection method  introducing the variables
\begin{equation}\label{Transf00}
x_r=\rho  \cos \theta  \sin \Phi  \sin \psi ,y_r=\rho  \cos \psi, z_r=\rho
\sin \theta  \sin\Phi
    \sin \psi,v_r=\rho  \cos \Phi \sin \psi,
\end{equation}
where $\rho=\frac{r}{\sqrt{1+r^2}},$ $r=\sqrt{x^2+y^2+z^2+v^2},$
$\theta,\Phi\in[0,2\pi],$
and $-\frac{\pi}{2}\leq \psi \leq \frac{\pi}{2}$
\cite{PoincareProj,Leon:2008de,Xu:2012jf,Leon2011}.

\begin{itemize}

\item{\bf Center manifold analysis for $F$}

Using the  Poincar\'e coordinates \eqref{Transf00}, the point $F$
transforms to the point with coordinates
$x_r=-\frac{\lambda_V }{2 \sqrt{3}},y_r=\frac{1}{2}
\sqrt{2-\frac{\lambda_V ^2}{3}},z_r=0,v_r=0$.
  Introducing the new variables
\begin{align}\label{varsF}
&u=v_r,\nonumber\\
&v_1= z_r,\nonumber\\
&
v_2= -\frac{1}{6} x_r \sqrt{6-\lambda_V ^2} \lambda_V +y_r
\left(2-\frac{\lambda_V
   ^2}{6}\right)-\frac{z_r \sqrt{1-\frac{\lambda_V ^2}{6}} \lambda_V ^4}{6
\left(\lambda_V ^2-2\right)}-\sqrt{2-\frac{\lambda_V ^2}{3}},\nonumber\\
&v_3= \frac{1}{6}
   x_r \sqrt{6-\lambda_V ^2} \lambda_V +\frac{1}{6} y_r \left(\lambda_V
^2-6\right)+\frac{1}{4} z_r \sqrt{1-\frac{\lambda_V ^2}{6}} \lambda_V
^2+\frac{1}{2}
   \sqrt{2-\frac{\lambda_V ^2}{3}},
\end{align}
we deduce that the local invariant center manifold of $F$ is given by the
approximated  graph
\begin{eqnarray}
&&\left\{(u,v_1,v_2,v_3)| v_1={\cal
O}\left(u^4\right),v_2=-\frac{1}{2} \sqrt{2-\frac{\lambda_V ^2}{3}}
u^2+{\cal O}\left(u^4\right),\right.\nonumber\\
&&\left.\ \ \ \ \ \ \ \ \ \ \ \ \ \ \ \ \ \ \ \ \ \ \ \ \ \ \ \ \ \ \ \ \
\ \,\ \ v_3=\frac{1}{4} \sqrt{2-\frac{\lambda_V ^2}{3}} u^2+{\cal
O}\left(u^4\right), |u|<\delta\right\},
\end{eqnarray}
 where $\delta>0$ is a suitably small positive real number and
${\cal O}(4)$ denotes terms of fourth order in the vector norm. Therefore,
the dynamics at the center manifold is given by
\begin{equation}
u'=\sqrt{2} u^2 \lambda_V,\; |u|<\delta,
\label{0centerF}
\end{equation}
with solution for $u(0)=u_{0}$ given by
\begin{equation}
u_1(\tau )=\frac{u_{0}}{1-\sqrt{2}u_{0} \lambda_V  \tau}.
\end{equation}
Since the origin is an inflection point of the potential
$U(u)=-\frac{\sqrt{2}}{3}{u^3\lambda_V}$ of the gradient-like equation
\eqref{0centerF}, it is implied that $F$ is a saddle point for
$\lambda_V^2<3$, only for perturbations in the direction of $v_r$-axis and
thus for perturbations in the $v$-axis.

\item{\bf Center manifold analysis for $F_0$}

Using the  Poincar\'e coordinates \eqref{Transf00}, the point $F_0$
transforms to the point with coordinates  $x_r=0,
y_r=\frac{\sqrt{2}}{2},z_r=0,v_r=0$. Proceeding similarly to point $F$
above and introducing the new variables
\begin{equation}\label{varsF0} u_1=v_r,\, v_1= x_r,\,
v_2=y_r-\frac{\sqrt{2}}{2},\,v_3=z,
\end{equation}
 we find  that the center manifold
is given by the graph
\begin{align}
&\left\{(u_1,v_1,v_2,v_3): v_1=-\frac{u_1^2}{2 \sqrt{2}}+{\cal O}(4),
v_2={\cal O}(4), v_3={\cal O}(4), |u_1|<\delta\right\},
\end{align}
where $\delta$ is a  suitably small constant and ${\cal O}(4)$ denotes
terms of fourth order in the vector norm. The dynamics on the center
manifold is governed by the equation
$u_1'={\cal O}(4)$, from which it follows that $F_0$ is indeed stable, but
for $n\neq0$ not asymptotically. This behavior is depicted in
Fig.~\ref{fig4}.
 \begin{figure}[ht!]
\begin{center}
\includegraphics[width=8cm,height=8cm,angle=0,clip=true]{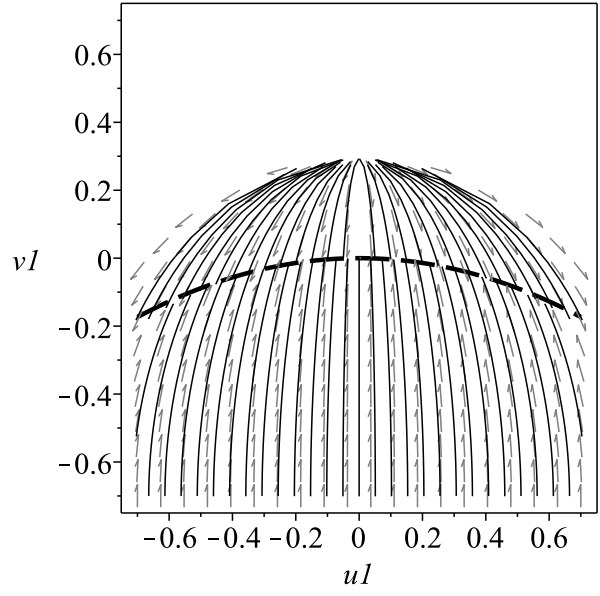}
\caption{{\it{ We consider the coordinate system \eqref{varsF0},
using $\lambda_V=0$ and $n$ arbitrary (for the numerics we
choose $n=1$ but different
$n$'s correspond to the same projection) . The dashed thick line represents
the
center manifold of $F_0$, which is stable. The orbits above and below that
line tend to it as time goes forward.}}}
\label{fig4}
\end{center}
\end{figure}

\item{\bf Center manifold analysis for $G$}

Using the  Poincar\'e coordinates \eqref{Transf00}, the point $G$
transforms to the point with coordinates
$x_r=-\frac{\sqrt{\frac{3}{2}}}{\sqrt{\lambda_V
^2+3}},y_r=\frac{\sqrt{\frac{3}{2}}}{\sqrt{\lambda_V
^2+3}},z_r=0,v_r=0$.
Introducing the new variables
{\small{
\begin{align}\label{varsG}
&u= v_r,\ \ \ \ \ \ \ \ \ \ \ \ \ \ \ \ \ \ \ \ \ \ \ \ \ \ \ \ \ \
\ \ \ \ \ \ \ \ \ \ \ \ \ \ \ \ \ \ \ \ \  \ \ \ \ \ \ \  \ \ \ \ \
\ \ \ \ \ \ \ \ \ \ \nonumber\\
&v_1= z_r,\nonumber
\end{align}
\begin{eqnarray}
&&v_2= -\frac{x_r \left(2 \lambda_V ^4-3 \lambda_V ^2+9\right)}{2 \lambda_V
 \sqrt{24-7 \lambda_V ^2} \left(\lambda_V
   ^2+3\right)}
-\frac{\lambda_V  \left(\sqrt{24-7 \lambda_V ^2}+\lambda_V
   \right)+6}{2 \lambda_V  \sqrt{16-\frac{14 \lambda_V ^2}{3}}
\sqrt{\lambda_V ^2+3}}\nonumber\\
&&  \ \ \ \ \ \,\  +\frac{y_r \left(-\lambda_V
^4+12 \lambda_V ^2+3 \sqrt{24-7 \lambda_V ^2} \lambda_V +\sqrt{24-7
\lambda_V ^2} \lambda_V ^3+9\right)}{2 \lambda_V  \sqrt{24-7
   \lambda_V ^2} \left(\lambda_V ^2+3\right)}\nonumber\\
&& \ \ \ \ \ \,  \ -\frac{3 \sqrt{\frac{3}{2}} z_r \left(\sqrt{24-7
\lambda_V ^2}+3 \lambda_V \right) \left(2 \lambda_V ^4-3 \lambda_V
   ^2+9\right)}{8 \lambda_V  \sqrt{24-7 \lambda_V ^2} \left(\lambda_V
^2+3\right) \left(2 \lambda_V ^2-3\right)},\nonumber
   \end{eqnarray}
\begin{eqnarray}
&&v_3= \frac{x_r \left(2 \lambda_V ^4-3 \lambda_V ^2+9\right)}{2 \lambda_V
   \sqrt{24-7 \lambda_V ^2} \left(\lambda_V ^2+3\right)}+\frac{\lambda_V
^2-\sqrt{24-7
   \lambda_V ^2} \lambda_V +6}{2 \lambda_V  \sqrt{16-\frac{14 \lambda_V
^2}{3}} \sqrt{\lambda_V ^2+3}}\ \ \nonumber\\
&& \ \ \ \ \ \  +\frac{y_r\left(\lambda_V ^4-12 \lambda_V ^2+3
\sqrt{24-7 \lambda_V ^2} \lambda_V +\sqrt{24-7 \lambda_V ^2} \lambda_V
   ^3-9\right)}{2 \lambda_V  \sqrt{24-7 \lambda_V ^2} \left(\lambda_V
^2+3\right)}\nonumber\\
&& \ \ \ \  \ \ -\frac{3 \sqrt{\frac{3}{2}} z_r \left(\sqrt{24-7
\lambda_V ^2}-3 \lambda_V \right)
   \left(2 \lambda_V ^4-3 \lambda_V ^2+9\right)}{8 \lambda_V  \sqrt{24-7
\lambda_V ^2} \left(\lambda_V ^2+3\right) \left(2 \lambda_V ^2-3\right)},
\end{eqnarray}}}
we find that the local invariant center manifold of $G$ is given by the
graph
{\small{
\begin{align}&\left\{(u,v_1,v_2,v_3)| v_1={\cal O}(4),\right. \nonumber\\ &
\left.  \ \
v_2=-\frac{\sqrt{\frac{3}{2}} u^2 \left[\lambda_V  \left(3 \lambda_V
^2-\sqrt{24-7 \lambda_V ^2} \lambda_V -15\right)-3 \sqrt{24-7 \lambda_V
^2}\right]}{2 \lambda_V
   \sqrt{\lambda_V ^2+3} \left(7 \lambda_V ^2-\sqrt{24-7 \lambda_V ^2}
\lambda_V -24\right)}+{\cal O}(4),\right. \nonumber\\
 & \left.
\ \ v_3=\frac{\sqrt{\frac{3}{2}} u^2 \left[\lambda_V  \left(-3 \lambda_V
^2-\sqrt{24-7
   \lambda_V ^2} \lambda_V +15\right)-3 \sqrt{24-7 \lambda_V ^2}\right]}{2
\lambda_V  \sqrt{\lambda_V ^2+3} \left(7 \lambda_V ^2+\sqrt{24-7 \lambda_V
^2} \lambda_V
   -24\right)}+{\cal O}(4), |u|<\delta\right\},
\end{align}}}
 with
$\delta>0$   a suitably small positive real number.
The dynamics at the center manifold is given by
\begin{equation}
u'=\frac{3 u^2 \sqrt{\lambda_V ^2+3}}{\lambda_V ^2},\;
|u|<\delta,\label{0centerG}
\end{equation}
with solution for $u(0)=u_0$ given by
\begin{equation}
u(\tau )=\frac{u_0 \lambda_V^2}{\lambda_V^2-3 u_0 \sqrt{\lambda_V^2+3} \tau
}.
\end{equation}
Since the origin is an inflection point of the potential  $U(u)=-\frac{u^3
\sqrt{\lambda_V ^2+3}}{\lambda_V ^2}$ of the gradient-like equation
\eqref{0centerG}, it follows that $G$ is a saddle point for either
$3<\lambda_V^2<\frac{24}{7}$ or $\lambda_V^2>\frac{24}{7}$, only for
perturbations
in the direction of $v_r$-axis and thus for perturbations in the $v$-axis.

\end{itemize}

\subsection{Stability of the critical points   at
infinity}\label{Poincare2}

We consider the Poincar\'e variables (\ref{Transf00})  (we
restrict the angle $\psi$ to this interval since the physical
region is given by $y>0$), and thus the points at
infinity
($r\rightarrow+\infty$) are those having $\rho\rightarrow 1$. Furthermore,
the physical phase space is given by
$$\left(x_r,y_r,z_r,v_r\right)\in [-1,1]\times [0,1]\times[-1,1]\times
[-1,1]$$ such that
\begin{equation}
\frac{x_r^2+y_r^2}{1-v_r^2-x_r^2-y_r^2-z_r^2}-\frac{x_r^2 z_r
\left(\sqrt{6} n v_r x_r-6 v_r^2-6
   x_r^2-6 y_r^2-6
z_r^2+6\right)}{\left(1-v_r^2-x_r^2-y_r^2-z_r^2\right)^{5/2}}\leq 1,
\end{equation}
and $v_r^2+x_r^2+y_r^2+z_r^2\leq 1.$

Performing the transformation \eqref{Transf00} on the system
(\ref{eqpowerx})-(\ref{eqpowerw}), and taking
the limit $\rho\rightarrow
1$, the leading terms in the resulting system
are
\begin{align}
&(1-\rho ^2) \rho'\rightarrow -\frac{1}{2} n (n+1) \cos ^2\theta  \sin
^2\Phi  \cos ^2\Phi  \sin ^4\psi  \left[2 \cos (2 \theta)  \sin
^2\Phi  \sin ^2\psi +\cos 2 \psi +1\right],
\label{inftyaa}\\
&(1-\rho ^2)^2 \theta'\rightarrow -\frac{1}{2} n (n+1) \sin \theta  \cos
^3\theta \, \sin ^2(2 \Phi)  \sin ^4\psi ,
\label{inftybb}\\
& (1-\rho ^2)^2 \Phi'\rightarrow n (n+1) \cos^2\theta  \sin ^2\Phi
\cos ^2\Phi  \sin ^5\psi  \cos \psi  \left[\cos (2 \theta ) \sin
^2\Phi -1\right],
\label{inftycc}\\
& (1-\rho ^2)^2 \psi'\rightarrow  n (n+1) \cos ^2\theta  \cos (2 \theta )
\sin ^3\Phi  \cos ^3\Phi \sin ^4\psi .\label{inftydd}
\end{align}
Since the equation for $\rho$ decouples, we only need to investigate the
subsystem of the angular variables. Thus, equating \eqref{inftybb},
\eqref{inftycc}
and \eqref{inftydd} to zero we obtain four classes of solutions:
\\
{\bf Class A}: \begin{eqnarray}
 &&x_r=0,\ y_r=\cos\psi, \ z_r=\pm \sin\Phi \sin \psi,\ v_r=\cos \Phi \sin
\psi, \label{PointsatinfinityAA}
\end{eqnarray} for some $\Phi, \psi.$
These points correspond to
$\theta=\pm\frac{\pi}{2}$. For these points,  $\Omega_{DE}$ of
(\ref{Omegas22powerlaw}) diverges as $\rho\rightarrow 1^-$, unless
$y_r=0$, thus the meaningful critical points at infinity
are those contained in the invariant circle $z_r^2+v_r^2=1$. Representative
critical points of this curve are the points
$$L^\pm:\left[x_r=0,y_r=0,z_r=\pm 1, v_r=0\right]$$ and
$$M^\pm:\left[x_r=0, y_r=0, z_r=0, v_r=\pm 1\right].$$
\\
{\bf Class B}: \begin{eqnarray}
&&x_r=0, \ y_r=\cos\psi, \  z_r=0,\ v_r=\pm \sin
\psi,\label{PointsatinfinityBB}
\end{eqnarray} for some $\psi.$ These points correspond to
$\Phi=0, \pi$, but since   $\Omega_{DE}$ of (\ref{Omegas22powerlaw})
diverges as $\rho\rightarrow 1^-$ unless $y_r=0$, all these points  are
unphysical apart from $$M^\pm:\left[x_r=0, y_r=0, z_r=0, v_r=\pm
1\right].$$
\\
{\bf Class C}: \begin{eqnarray}
 &&x_r=0, \ y_r=1, \ z_r=0, \ v_r=0.\label{PointsatinfinityCC}
\end{eqnarray} These points correspond to $\psi=0$. They  lead to a
divergence in $\Omega_{DE}$
in (\ref{Omegas22powerlaw}) and thus they are unphysical.
\\
{\bf Class D}: \begin{eqnarray}
 &&x_r=\pm \cos \theta \sin \psi, \ y_r=\cos \psi, \ z_r=\pm \sin \theta
\sin \psi, \ v_r=0.\label{PointsatinfinityDD}
\end{eqnarray}
These points correspond to $\Phi=\pm\frac{\pi}{2}$,
 but since   $\Omega_{DE}$ of (\ref{Omegas22powerlaw})
diverges unless $x_r=y_r=0$, all these points  are
unphysical apart from  $$L^\pm:\left[x_r=0,y_r=0,z_r=\pm 1,
v_r=0\right].$$

>From the above analysis we deduce that for $n\neq0$  the singular points at
infinity satisfy $x_r=0$ (we mention that $x_r=0$
is not an invariant set for the flow unless $y_r= 0$ or
$\lambda=0$) that is points $M^\pm$, or  $v_r=0$ that is points $L^\pm$.

Analyzing the dynamical system for $x_r,y_r,z_r,v_r$  around  $M^\pm$,  we find
that
the eigenvalues of the linearization are
$\left(0,\frac{3}{2}, -\frac{3}{2}, \frac{9}{2}\right)$, and thus, although
non-hyperbolic, points
$M^\pm$ behave as saddle points for the Poincar\'e vector field.

In order to calculate the eigenvalues of the linearization of the
dynamical system for $x_r,y_r,z_r,v_r$  around $L^\pm$, we proceed as
follows: having calculated the linear perturbation matrix $\mathbf{Q}$ we
evaluate it at $y_r=v_r=0$, then we take the limit  $x_r\rightarrow 0$ and
finally we incorporate the lateral limit $z_r\rightarrow 1^-$ (that is
from below) for $L^+$ or $z_r\rightarrow -1^+$ (that is from above) for
$L^-$. Calculating the eigenvalues of the resulting   matrix we
obtain that the eigenvalues of the linearization for both $L^\pm$ are given
by
$\left(3, \frac{15}{4}, \frac{9}{2},\frac{9}{4}\right)$. Thus, we deduce
that points $L^\pm$ are unstable.

\end{appendix}

\end{document}